\documentclass{aa}  

\usepackage{graphicx}
\usepackage[varg]{txfonts}
\usepackage{color}
\usepackage[switch]{lineno} 
\usepackage{longtable}
\usepackage{natbib}
\usepackage{hyperref}
\bibpunct{(}{)}{;}{a}{}{,} 
\newcommand{\fer}{{\it Fermi}}
\newcommand{\vv}{V407\,Cyg}
\newcommand{\kms}{km\,s$^{-1}$}

\begin{document}

\title{Very Long Baseline Interferometry imaging of the advancing ejecta in the first gamma-ray nova V407\,Cyg}

\author{M.~Giroletti\inst{1}\fnmsep\thanks{Email: giroletti@ira.inaf.it}, 
  U.~Munari\inst{2},
  E.~K\"ording\inst{3}, 
  A.~Mioduszewski\inst{4},
  J.~Sokoloski\inst{5,6},
  C.\,C.~Cheung\inst{7},
  S.~Corbel\inst{8,9},
  F.~Schinzel\inst{10}\fnmsep\thanks{is also an adjunct professor at the University of New Mexico},
  K.~Sokolovsky\inst{11,12,13},
  T.\,J.~O'Brien\inst{14}
}

\institute{INAF Istituto di Radioastronomia, via Gobetti 101, 40129 Bologna, Italy \and
INAF Astronomical Observatory of Padova, 36012 Asiago (VI), Italy \and
Department of Astrophysics/IMAPP, Radboud University Nijmegen, 6500 GL Nijmegen, the Netherlands \and
National Radio Astronomy Observatory, Array Operations Center, 1003 Lopezville Road, Socorro, NM 87801, USA \and 
Columbia Astrophysics Laboratory, Columbia University, New York, NY 10027, USA \and
LSST Corproation, 933 North Cherry Avenue,
Tucson, AZ 85721, USA \and
Space Science Division, Naval Research Laboratory, Washington, DC 20375, USA \and
Laboratoire AIM (CEA/IRFU - CNRS/INSU - Universit\'e Paris Diderot), CEA DSM/IRFU/SAp, F-91191 Gif-sur-Yvette, France \and
Station de Radioastronomie de Nan\c{c}ay, Observatoire de Paris, CNRS/INSU, USR 704 - Univ. Orl\'eans, OSUC, 18330 Nan\c{c}ay, France \and
National Radio Astronomy Observatory, P.O. Box O, Socorro, NM, 87801, USA \and
Department of Physics and Astronomy, Michigan State University, 567 Wilson Rd, East Lansing, MI 48824, USA \and
Astro Space Center, Lebedev Physical Inst. RAS, Profsoyuznaya 84/32, 117997 Moscow, Russia \and
Sternberg Astronomical Institute, Moscow University, Universitetsky 13, 119991 Moscow, Russia \and
Jodrell Bank Centre for Astrophysics, Alan Turing Building, University of Manchester, Manchester M13 9PL, UK }

\date{Received ; accepted }

  \abstract
{In 2010 March, the Large Area Telescope on board \fer\ revealed a transient gamma-ray source, positionally coincident with the optical nova in the symbiotic binary, \vv. This event marked the first discovery of gamma-ray emission from a nova.}
{We aimed to obtain resolved radio imaging of the material involved in the nova event; to determine the ejecta geometry and advance velocity directly in the image plane; to constrain the physical conditions of the system.}
{We observed the source with the European VLBI (Very Long Baseline Interferometry) Network in real time mode, at 1.6 and 5 GHz, and the Very Long Baseline Array at 1.6, 5, and 8.4 GHz. In total, we observed the source over 16 epochs, starting 20 days after the optical discovery and continuing for over 6 months.}
{Milliarcsecond scale radio emission is detected in 10/16 epochs of
observations.  The source is initially very dim but it later shows a
substantial increase in brightness and a resolved shell-like structure 40 to 90 days after the optical event.  The shell has a projected
elliptical shape and is asymmetric in brightness and spectral index, being
brighter and characterised by a rising spectrum at the south-eastern
edge.  We determine a projected
expansion velocity of $\sim$3\,500 \kms\ in the initial phase (for an
adopted 2.7 kpc distance), and $\sim$2\,100 \kms\ between day 20 and 91. 
We also found an emitting feature about 350 mas
(940 AU) to the north-west, advancing at a projected velocity
of $\sim$700 \kms\ along the polar axis of the binary. The total flux
density in the VLBI images is significantly lower than that previously
reported at similar epochs and over much wider angular scales with the 
VLA.}
{Optical spectra convincingly demonstrated that in 2010 we were viewing \vv\ along the equatorial plane and from behind the Mira.  Our radio observations image the bipolar flow of the ejecta perpendicular to the orbital plane, where deceleration is much lower than through the equatorial plane probed by the truncated profile of optical emission lines.  The separated polar knot at 350 mas and the bipolar flow strictly resemble the similar arrangement seen in Hen 2-104, another symbiotic Mira seen equator-on that went through a large outburst $\sim$5700 yrs ago.  The observed $\sim$700 \kms\ expansion constrains the launch-date of the polar knot around 2004, during the accretion-fed active phase preceding the 2010 nova outburst.  
}


  \authorrunning{M.\ Giroletti et al.}
  \titlerunning{VLBI imaging of $\gamma$-ray nova V407 Cyg 2010}

   \maketitle

\begin{table*}
   \small
   \centering
   \caption{
Log of EVN observations.}
   \label{t.logevn}
   \begin{tabular}{lclcccc}
   \hline
   \hline
Date & Time &  &  &  & Half Peak Beam &  \\
in 2010 & since nova & Participating stations & Freq. & Duration & Width (HPBW) & $1\sigma$ rms \\
 & (days) & & (GHz) & (hrs) & (mas $\times$ mas, $^\circ$) & ($\mu$Jy beam$^{-1}$) \\
\hline
March 30	& 20  & Ef, Jb2, Mc, On, Sh, Wb, Ys, Cm, Kn 	& 5   & 7.5  & $12.1\times10.2,77$ & 29 \\
April 23	& 44  & Ef, Jb1, Mc, On, Tr, Wb, Cm, Kn     	& 1.6 & 11.5 &  $28\times 26, -45$ & 18 \\
May 19	& 70  & Ef, Jb2, Mc, On, Tr, Wb, Ys, Cm, Kn 	& 5   & 11  & $8.3\times7.6, 56$ &   55 \\
June 9       & 91  & Ef, Jb1, Mc, On, Tr, Wb, Cm, Da      	& 1.6 & 11.5 & $27 \times 25, 58$ & 26 \\
Sept.\ 8	& 182 & Ef, Jb2, Mc, On, Tr, Wb, Ys         	& 5   & 9.5   & $10.0\times9.3,-72$ & 26 \\
Sept.\ 29	& 203 & Ef, Jb2, Mc, On, Tr, Wb, Cm         	& 1.6 & 12  & $ 22\times22,46$ & 60 \\
\hline
   \end{tabular} 
\tablefoot{Station codes: Cm – Cambridge (32 m), Da – Darnhall (25 m), Ef – Effelsberg (100 m), Jb1 – Jodrell Bank (Lovell Telescope, 76 m), Jb2 – Jodrell Bank (MarkII telescope, 32\,m$ \times  25$ m), Kn - Knockin (25 m), Mc – Medicina (32 m), On – Onsala (20 m), Sh – Shanghai (25 m), Tr – Torun (32 m), Ys – Yebes (40 m), Wb – Westerbork Synthesis Radio Telescope (WSRT, 12 dishes $ \times  25$ m).}
\end{table*}

\begin{table*}
   \small
   \centering
   \caption{
Log of VLBA observations.}
   \label{t.logvlba}
   \begin{tabular}{lcccccccc}
   \hline
   \hline

Date & Time & Total & \multicolumn{2}{c}{1.6 GHz data} & \multicolumn{2}{c}{5 GHz data} & \multicolumn{2}{c}{8.4 GHz data} \\ 
in 2010 & since nova & duration & HPBW &  $1\sigma$ rms & HPBW &  $1\sigma$ rms & HPBW &  $1\sigma$ rms \\ 
 & (days) & (hrs) & (mas $\times$ mas, $^\circ$) & ($\mu$Jy beam$^{-1}$) & (mas $\times$ mas, $^\circ$) & ($\mu$Jy beam$^{-1}$) & (mas $\times$ mas, $^\circ$) & ($\mu$Jy beam$^{-1}$) \\
\hline
March 28\tablefootmark{a} & 18 & 3 & \multicolumn{2}{c}{\dots}  & $5.1\times4.3, -60$ & 90 & \multicolumn{2}{c}{\dots} \\
April 2 				& 23 & 4 & $14.1\times5.4, -13$ 	& 260 & $5.4\times4.1, -43$ 	& 110 & $5.0\times4.0, -48$ & 75 \\
April 6\tablefootmark{b} 	& 27 & 4 & $14.8\times10.6, -18$ 	& 160 & $4.7\times3.7, -32$ 	& 120 & \multicolumn{2}{c}{\dots} \\
April 10 				& 31 & 4 & $13.5\times9.1, -69$ 	& 85 & $5.0\times4.3, -55$ 	& 95 & \multicolumn{2}{c}{\dots} \\
April 15\tablefootmark{c} 	& 36 & 4 & $14.0\times10.8, -11$ 	& 100 & $4.9\times4.1, -34$ 	& 85 & \multicolumn{2}{c}{\dots} \\
April 21\tablefootmark{d} 	& 42 & 4 & $15.1\times12.8, -46$ 	& 130 & $5.3\times4.0, -46$ 	& 120 & \multicolumn{2}{c}{\dots} \\
April 29\tablefootmark{e} 	& 50 & 4 & $14.6\times13.3, 13$ 	& 130 & $5.2\times4.2, -19$ 	& 100 & \multicolumn{2}{c}{\dots} \\
May 4\tablefootmark{f} 	& 55 & 4 & $15.2\times13.7,41$ 	& 95 & $5.0\times4.1, -38$ 	& 90 & \multicolumn{2}{c}{\dots} \\
May 14\tablefootmark{g} 	& 65 & 4 & $13.4\times9.0, 19$ 	& 140 & $5.0\times3.7, 10$ 	& 110 & \multicolumn{2}{c}{\dots} \\
May 21\tablefootmark{h} 	& 72 & 4 & $13.7\times10.2,-10$ 	& 85 & $5.0\times4.0, -22$ 	& 90 & \multicolumn{2}{c}{\dots} \\
\hline
   \end{tabular} 
\tablefoot{All dishes have the same 25\,m diameter.
\tablefoottext{a}{No Mk. Phase calibrator not detected at 1.6 GHz.}
\tablefoottext{b}{No Fd, Kp, Mk.}
\tablefoottext{c}{No useful data from Mk at 1.6 GHz.}
\tablefoottext{d}{No Fd, Sc; Mk did not give useful data at 1.6 GHz.}
\tablefoottext{e,f}{Mk, Sc did not give useful data at 1.6 GHz.}
\tablefoottext{g}{Fd no good data at either 1.6 or 5 GHz; Hn, Mk, Sc very few useful data at 1.6 GHz.}
\tablefoottext{h}{Mk, Sc very few useful data at 1.6 GHz.}
}
\end{table*}

\section{Introduction}

\vv\ was already a remarkable symbiotic binary before the spectacular nova
outburst it underwent in 2010 \citep[for a recent review of symbiotic stars and their outburst mechanisms, see][]{Munari2019}.  The companion to the accreting white dwarf (WD) is a Mira with an exceedingly long pulsation period of 745 days, the longest known among symbiotic stars (at the time of writing, among the $\sim$20,000 field Mira catalogued in VSX\footnote{Variable Star indeX, \url{https://www.aavso.org/vsx/}}, only 0.1\% has a pulsation period longer than \vv).  Miras with such long pulsation periods are generally OH/IR sources, enshrouded in a very thick dust envelope which prevents direct observation of the central star at optical wavelengths.  V407 Cyg is instead well visible down to the short wavelengths probed by ultraviolet spectra collected by the IUE satellite in 1982 and 1991.  The factors inhibiting the formation of a massive dust cocoon also in \vv\ probably reside in the ionisation action exerted by the accreting WD and the sweeping of the circumstellar space by the ejecta during nova outbursts.  A slow modulation of the Mira brightness led \citet{Munari1990} to speculate about a $\geq$43 yr orbital period, recently upwarded to $\geq$100 yr by the long-term radial velocity monitoring performed by \citet{Hinkle2013}.  \vv\ was first noticed when it underwent an eruption in 1936 \citep{Hoffmeister1949}.  This active phase lasted for about 3 years, while a second and stronger one was discovered in 1994 \citep{Munari1994}, peaked in brightness in 1998 \citep{Kolotilov1998,Kolotilov2003} and was still ongoing at the time of the nova outburst in 2010 \citep{Munari2011}.  Both 1936 and 1998 eruptions were of the accretion-fed type, and we will show in this paper how their products still had a role in the 2010 event.  

The highlight of its career is the violent nova outburst it underwent in March 2010, granting \vv\ a special place among novae in being the first one detected in GeV $\gamma$-rays \citep{Abdo2010}. At that time, the prevailing opinion was that novae would have to wait the launch of future, more sensitive satellites to record the MeV $\gamma$-rays produced by $e^{+}-e^{-}$ annihilation and by the decay of unstable isotopes (as $^{13}$N, $^{18}$F, $^{7}$Be and $^{22}$Na) synthesised during the brief initial thermonuclear runaway \citep[e.g.][]{Hernanz2004};  particle acceleration up to TeV energies had however been proposed based on the properties of the 2006 outburst of RS Oph \citep{Tatischeff2007}.

The 2010 outburst of V407 Cyg was independently discovered at unfiltered 7.4 mag on March 10.797 UT by K.\ Nishiyama and F.\  Kabashima, and by other amateur astronomers in the following nights (see \citealt{Nishiyama2010} for a report on the initial detections). Spectroscopic confirmation of a nova erupting within a symbiotic binary was soon provided by \citet{Munari2010}, followed a few days later by early near-IR spectroscopy by \citet{Joshi2010}.  It took however a whole week to raise the interest of a much wider community, triggering observations over the full wavelength range, radio included:  routine \fer-LAT processing of the all-sky data led \citet{Cheung2010} to announce, a week later, that they had discovered a new transient gamma-ray source in the Galactic plane, Fermi J2102+4542, with $E>100$\,MeV flux of $(1.0\pm0.3) \times 10^{-6}$\,ph\,cm$^{-2}$\,s$^{-1}$ on March 13 and $(1.4\pm0.4) \times 10^{-6}$\,ph\,cm$^{-2}$\,s$^{-1}$ on March 14, and this new source was labelled as {\it possibly} associated with the eruption of \vv. Following the latter, several other novae have been detected in $\gamma$-rays with Fermi \citep[e.g.][]{Ackermann2014,Cheung2016}.

\begin{figure*}
\sidecaption
  \includegraphics[width=12cm]{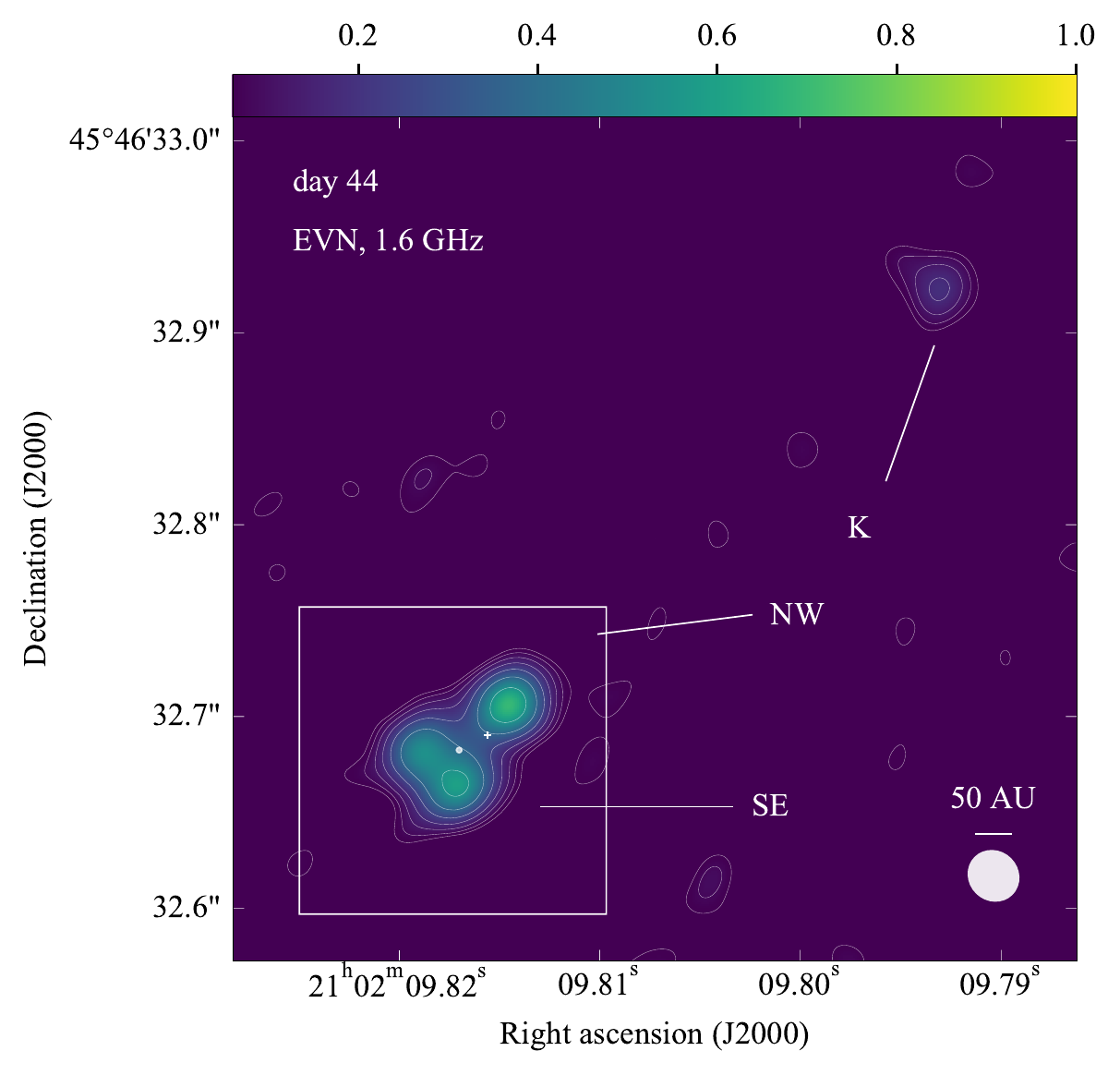}
     \caption{EVN image of \vv\ at 1.6 GHz on day 44. Beam size (shown by the ellipse on the lower-right corner) and noise levels are given in Table \ref{t.logevn}; contours start at $3\sigma$ and increase by steps of $\sqrt 2$; the top wedge shows the color scale between 0.06 and 1.0 mJy beam$^{-1}$. The white cross represents the position, and its uncertainty (magnified by a factor of 10 for visibility), of \vv\ in Gaia DR2 brought back to $t=0$ by application of proper motions; the white dot indicates the position of the assumed centre of expansion (see text for details).  The white box represents the inner region shown in Fig.~\ref{f.day70}.}
     \label{f.day44}
\end{figure*}

Radio observations of V407 Cyg commenced on March 22 with a number of low angular resolution instruments \citep{Nestoras2010, Gawronski2010, Pooley2010, Bower2010} that reported positive detection of the source; on the contrary, historical pre-outburst observations known at the time had not revealed this system as a radio source \citep{Wendker1995, Ivison1995} and only a later re-analysis of old Very Large Array (VLA) data for 1993 May 14 provided a clear detection of 1.18$\pm$0.07 mJy at 8.4 GHz \citep{Chomiuk2012}. At this epoch \vv\ was however already rising from quiescence toward the peak of the 1998 accretion-fed outburst.

In the 2006 (i.e.\ pre-\fer) nova outburst of another symbiotic system, RS\,Oph, Very Long Baseline Interferometry (VLBI) observations had revealed important details about the physics, such as the non thermal nature of the emission and the asymmetric jet-like ejection \citep[e.g.][]{O'Brien2006,Sokoloski2008,Rupen2008}. Because of the unique detection of gamma-ray emission from \vv, we immediately proposed to observe it with VLBI. We adopted a combined approach exploiting the advantages
offered by the European VLBI Network (EVN) and the Very Long Baseline Array (VLBA). The  latter offers frequency agility, scheduling flexibility, and denser time coverage. The former provides enhanced sensitivity as a result of a high data rate acquisition and the inclusion of the large apertures on short baselines; most importantly, it provides real-time correlation thanks to the so-called e-VLBI technique in which data are acquired and transmitted in real-time from the individual stations to the EVN data correlator. Indeed, the prompt discovery of compact emission in the source with the EVN \citep{Giroletti2010} was the basis of
the whole observational campaign presented in this paper. We will show how the combination of tight constraints from profiles of optical emission lines \citep{Shore2011,Shore2012,Munari2011} and the multi-epoch VLBI radio images here presented provide a robust 3D geometric and kinematical modelling of \vv\ and its circumstellar space.

Throughout the paper, we adopt a distance of 2.7\,kpc for
V407\,Cyg \citep{Munari1990}, corresponding to a linear scale of $4.0\times10^{13}$\,cm\,mas$^{-1}$ (=2.7 AU mas$^{-1}$). This value is intermediate between the $\geq$3.0 kpc adopted by \citet{Chomiuk2012} on the basis of interstellar absorption lines seen in optical spectra of V407 Cyg and the 2.3 kpc deriving from application of the Period-Luminosity relation for Mira calibrated by \citet{Whitelock2008}. Unfortunately, the parallax quoted in Gaia DR2 is useless in view of the twice larger value reported for the associated error. However, the great accuracy of Gaia astrometric position \citep{Gaia2018} is useful to place the star in comparison to the radio ejecta. We define the radio spectral index $\alpha$ such that $S(\nu)\propto\nu^{\alpha}$. We refer dates to the optical event, such that day 0 = 2010 March 10.0 UT. Note that on this scale the first discovery of the eruption on optical images by Nishiyama and Kabashima occurred on day=+0.797.

 \begin{figure*}
\sidecaption
  \includegraphics[width=12cm]{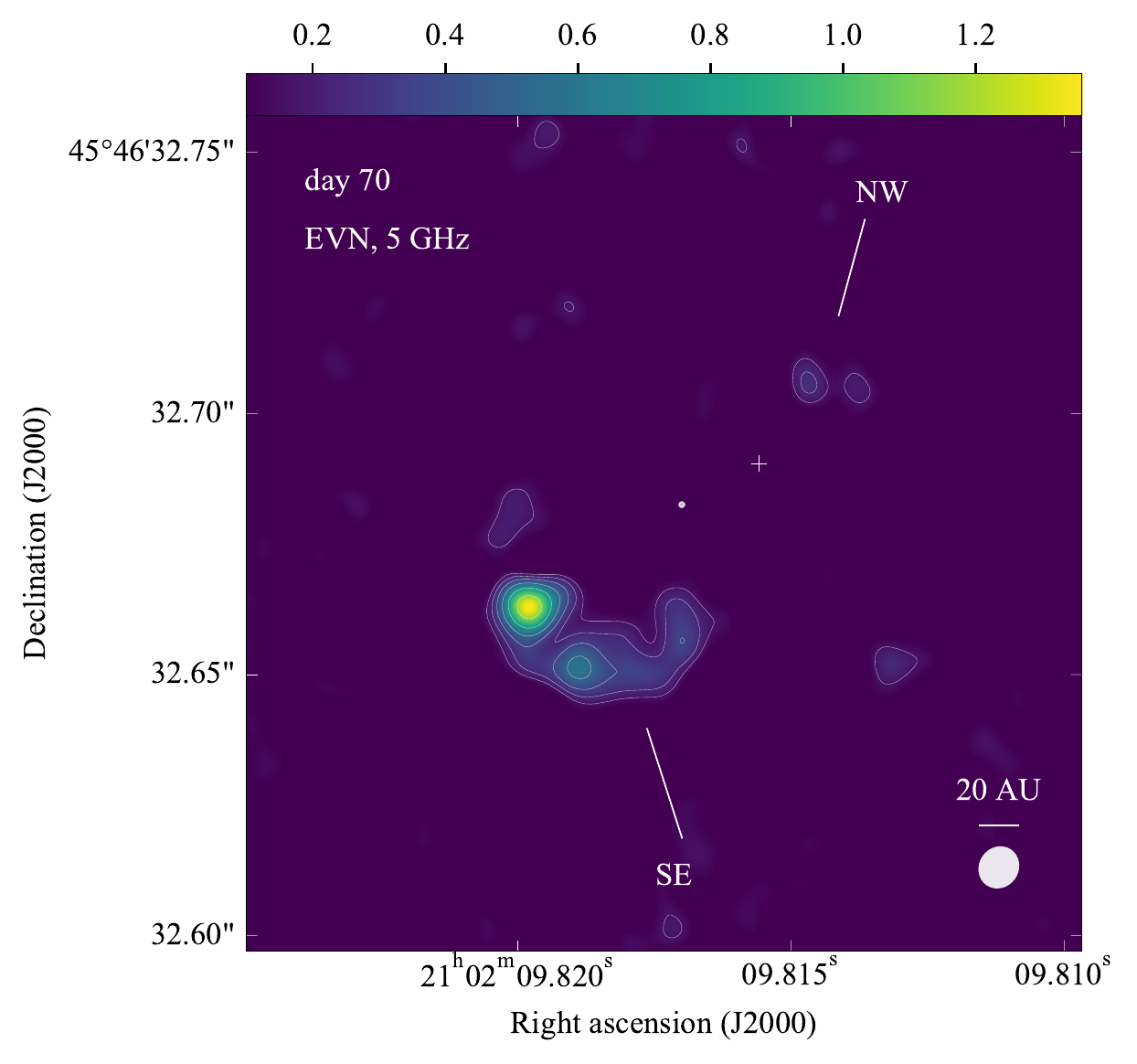}
     \caption{EVN image of \vv\ at 5 GHz on day 70. Beam size (shown by the ellipse on the lower-right corner) and noise levels are given in Table \ref{t.logevn}; contours start at $3\sigma$ and increase by steps of $\sqrt 2$; the top wedge shows the color scale between 0.1 and 1.36 mJy beam$^{-1}$. The white cross represents the position, and its uncertainty (magnified by a factor of 5 for visibility), of \vv\ in Gaia DR2 brought back to $t=0$ by application of proper motions; the white dot indicates the position of the assumed centre of expansion (see text for details).  The total size of this image corresponds to the white inset indicated in the larger panels of Figs.~\ref{f.day44} and~\ref{f.day91}.}
     \label{f.day70}
\end{figure*}

 \begin{figure*}
\sidecaption
  \includegraphics[width=12cm]{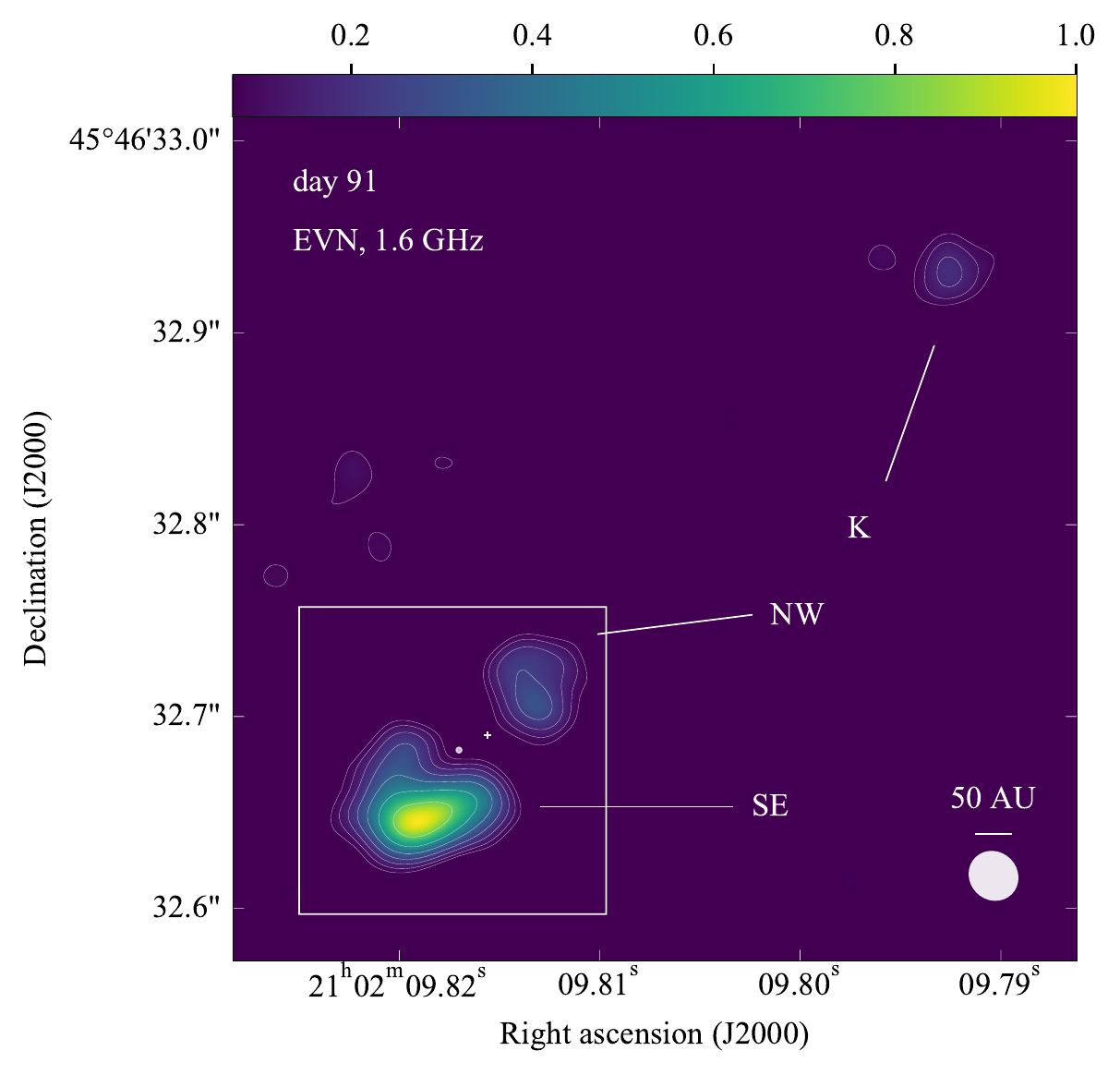}
     \caption{EVN image of \vv\ at 1.6 GHz on day 91. Beam size (shown by the ellipse on the lower-right corner) and noise levels are given in Table \ref{t.logevn}; contours start at $3\sigma$ and increase by steps of $\sqrt 2$; the top wedge shows the color scale between 0.06 and 1.0 mJy beam$^{-1}$. The white cross represents the position, and its uncertainty (magnified by a factor of 10 for visibility), of \vv\ in Gaia DR2 brought back to $t=0$ by application of proper motions; the white dot indicates the position of the assumed centre of expansion (see text for details).  The white box represents the inner region shown in Fig.~\ref{f.day70}.}
     \label{f.day91}
\end{figure*}

\section{Observations\label{s.2}}

\subsection{EVN observations}

We observed V407 Cyg six times with the EVN between 2010 March 30 and September 29, i.e.\ between 20 days and 6.7 months after the optical event. The observations were carried out alternating runs at 5 GHz ($\lambda= 6$ cm), on March 30 ($t=20$ d), May 19 ($t=70$ d), and September 8 ($t=182$ d) and at 1.6  GHz ($\lambda= 18$ cm), on April 23 ($t=44$ d), June 9 ($t=91$ d), and September 29 ($t=203$ d). A bandwidth of 1 Gbps was sustained by most stations, corresponding to eight 16 MHz wide sub-bands, 2 polarisations, and 2-bit sampling. A summary of the details about the participating telescopes and the observation length is given in Table~\ref{t.logevn}. The Table note also reports the diameter of each element participating in the array.

We observed in phase reference mode, using the source J2102+4702 as a phase calibrator at all epochs. The offset between the target and the phase calibrator is $1.26^\circ$, small enough to warrant a proper transfer of the phase solutions. The two sources were observed with the repetition of 200\,s on-target integrations bracketed by 80\,s scans of the calibrator. Correlation was performed in real time at the Joint Institute for VLBI in Europe (JIVE); the JIVE pipeline was also used to carry out a priori amplitude calibration, automated flagging, and fringe fitting with Astronomical Image Processing System (AIPS) tasks. We edited the final visibility data and produced clean images in Difmap \citep{Shepherd1994}. Owing to the low flux density in the source, we did not perform self-calibration.

The beam size (HPBW in Table~1) and image noise levels vary according to length of the observing run and weighting schemes; natural weights were used for the final images, whose parameters are reported in Table~\ref{t.logevn}. The noise levels are low (between 18 and 60 $\mu$Jy\,beam$^{-1}$), within a factor of a few times the predicted thermal noise of the receivers. The final epoch has somewhat worse noise level, due to the limited participation of the 100m Effelsberg radio telescope which contributed data for only about 4 hours.

\subsection{VLBA observations}

We observed the source with the VLBA 10 times, starting as soon as 2010 March 28 ($t= 18$ d) and ending on 2010 May 21 ($t= 72$ d). Exploiting the frequency agility of the VLBA, we observed each epoch at both 1.6 and 5 GHz (the same frequencies of EVN observations); on April 2, we also observed at 8.4 GHz ($\lambda= 3.6$ cm). At each frequency, we observed with four 8 MHz sub-bands, dual polarisation, and 2-bit sampling (for a total data rate of 256 Mbps). Each observation lasted 4 hours in total, except for the first epoch, which lasted 3 hours.

We used the same calibrator (J2102+4702) as a phase reference source as in the EVN observations, with 45\,s calibration scans bracketing 120\,s scans on source; the faster slew rate of the 25-m VLBA dishes allowed us to sustain a shorter cycle for calibration than with the EVN. Correlation was carried out using the DiFX correlator in Socorro \citep{Deller2011}; post correlation analysis was carried out in AIPS following standard procedures. We produced images with different weighting schemes; in general, the use of natural weights and of a Gaussian taper improved the image quality. We thus report in Table~\ref{t.logvlba} the log of our VLBA observations and the list of the final image parameters obtained with {\tt ROBUST = 5} and, for the 5 and 8.4 GHz observations, a circular taper of 40 M$\lambda$.  We also note in the Table some failures that occurred during the observations; in particular, the phase calibrator is faint and resolved out at 1.6 GHz on the longest VLBA baselines, which resulted in the lack of useful data at this frequency for the first epoch on the entire array, and for baselines to Mauna Kea (MK) and Saint Croix (SC) in several later ones.

\section{Results\label{s.3}}

\begin{table*}
   \centering
   \caption{
\label{t.modelfit}Results of modelfit. The reference position is at R.A.\ 21h 02m 09.81700s,  Dec.\ $+45^\circ$ 46$^\prime$ 32$^{\prime\prime}$.68253 }
   \begin{tabular}{cccccccccc}
   \hline
   \hline
Time since nova & Array & Freq. & $r$ & $\theta$ & $S_\nu$ & $a$ & $b$ & $\phi$ & region \\
(days) & & (GHz)  & (mas) & ($^\circ$) & (mJy) & (mas)  & (mas) & ($^\circ$) \\
(1) & (2) & (3) & (4) & (5) & (6) & (7) & (8) & (9) & (10) \\
\hline
 20 & EVN & 5 & 9.3	& 171.7	& 0.2	& 7.7	& 1.6	& $-59.8$	& SE \\
 31 & VLBA & 5   & 13.5	& 100.1	& 1.7	 & 9.9	& 3.0 & $-23.8$ & SE \\
 36 & VLBA & 5   & 16.3	& 104.9	& 2.2	& 7.2 &	3.3	& $-25.9$  & SE \\
 44 & EVN & 1.6 & 19.0	& 90.6	& 0.8	& 19.3	& 19.3	& 27.2	& SE \\
 & &               	& 19.1	&$-177.1$	& 0.7	& 23.6	& 7.4	& $-46.8$	& SE \\
 & &               	& 33.8	& $-46.1$	& 0.9	& 20.2	& 11.0	& $-71.2$	& NW \\
 & &               	& 349.0	& $-45.9$	& 0.3	& 31.8	& 15.2	& 55.4	& K \\
 55 	& VLBA 	& 1.6	& 35.7	& 159.6	& 2.5	& 94.8	&8.9	& 71.2	& SE \\
 			&		&	& 45.7	& $-43.3$	& 1.3	& 34.3	&5.0	& $-53.9$ & NW \\
 70 & EVN & 5 & 33.8	& 105.9	& 1.8	& 29.9	& 10.5	& $-5.0$	& SE \\
 & &               	& 35.0	& 124.2	& 1.7	& 4.0	& 4.0	& 0.0	& SE \\
 & &               	& 36.6	& 150.4	& 2.3	& 17.7	& 7.0	& 89.7	& SE \\
 & &               	& 22.0	& $-177.4$& 1.3	& 14.2	& 14.2	& $-30.7$	& SE \\
 & &               	& 36.5	& $-53.5$	& 1.2	& 16.7	& 8.8	& 86.0	& NW \\
 72 & VLBA	& 1.6	& 39.0	& 135.1	& 0.6	& 10.5	& 10.5	& 0	& SE \\
		 	&		&	& 35.0	& $-63.1$	& 0.4	& 5.0		& 5.0		& 0	& NW \\
 72 & VLBA	& 5	& 36.2	& 123.5	& 1.7	& 3.6		& 1.9		& $-41.0$ & SE \\
		 	&		&	& 36.7	& 155.3	& 1.3	& 11.0	& 3.2		& $-55.6$ & SE \\
 91 & EVN & 1.6 & 28.7	& 88.1	& 0.3	& 10.2	& 10.2	& 0.0	& SE \\
 & &               	& 50.7	& 124.7	& 0.6	& 29.0	& 29.0	& 3.3	& SE \\
 & &               	& 43.2	& 149.3	& 1.2	& 15.2	& 15.2	& $-2.1$	& SE \\
 & &               	& 28.7	& $-169.5$& 0.9	& 17.6	& 17.6	& 40.9	& SE \\
 & &               	& 49.7	& $-49.5$	& 1.0	& 40.5	& 29.1	& $-11.9$	& NW \\
 & &               	& 357.7	& $-45.4$	& 0.4	& 31.7	& 19.0	& $-45.0$	& K \\
182 & EVN & 5 & 68.2	& 165.8	& 0.4	& 13.7	& 13.7	& 21.6	& SE \\
 & &               	& 49.9	& $-32.0$	& 0.1	& 6.0	& 6.0	& 15.1	& NW \\
203 & EVN & 1.6 & 55.1	& 146.8	& 2.8	& 135.7	& 49.1	& 79.5	& SE \\
 & &               	& 74.3	& $-67.4$	& 0.9	& 69.9	& 32.1	& $-29.3$	& NW \\
 & &               	& 372.2	& $-46.5$	& 0.2	& 9.7	& 9.7	& 12.4	& K \\
\hline
\end{tabular}
\tablefoot{
Cols.~(1, 2, 3):\ observation epoch, array, and frequency: Cols.~(4, 5):\  polar coordinates $r$ and $\theta$ with respect to the position R.A. 21h 02m 09.81700s,  Dec.\ $+45^\circ$ 46$^\prime$ 32$^{\prime\prime}$.68253 (for the determination of the reference point, see main text); Col.~(6):\ flux density $S_\nu$; Cols.~(7, 8, 9):\  major and minor axis $a$ and $b$ and position angle $\phi$ of the component; Col.~(10):\ the location within the source structure.
}
\end{table*}

We detect radio emission in our VLBI data for 11 datasets obtained on 10 epochs: all the six observations with the EVN, and four of the VLBA epochs (three of which at one single frequency and one at both 1.6 and 5 GHz).  In Figs.~\ref{f.day44}, \ref{f.day70}, \ref{f.day91}, we show the three images where we detect the highest flux density, all obtained with the EVN, at days 44, 70, and 91.  The whole set of images obtained during the campaign is shown in Appendix~\ref{s.appendixVLBI}: in Fig.~\ref{f.vlbi_in}, a composite of the inner 160 mas $\times$ 160 mas regions for all epochs with a detection; in Fig.~\ref{f.vlbi_out}, a set of wider 420 mas $\times$ 420 mas fields for the three epochs (days 44, 91, and 203) in which emission is detected also at larger distance.  In all the images, we show the position of \vv\ at the time of the nova, as obtained by extrapolating the coordinates reported in Gaia DR2 \citep{Gaia2018} with the application of the given proper motions. At the time of the Gaia observations, the light from \vv\ was completely dominated by the Mira (cf. Munari et al., in prep.) and therefore the astrometric position listed in DR2 pertains to the Mira more than to binary system as a whole; further details about the extrapolation of the Gaia data are given in Appendix~\ref{s.appendixGaia}.  We also show a bar corresponding to a reference linear scale, of size comparable to the orbit separation (for illustration purposes, we show a 50 AU ruler in the 1.6 GHz images and a 20 AU one in the 5 GHz image); see also Sect.~\ref{s.discussion-1} for a discussion of the linear scales involved in the system.

The images reveal a clear evolution in structure and total flux density. The first detection is achieved on day 20 with the EVN at 5 GHz.  At this epoch (shown in the first panel of Fig.~\ref{f.vlbi_in}), we reveal a component significant at the $\sim 6 \sigma$ confidence level, with a  peak brightness of $\sim0.16$\,mJy\,beam$^{-1}$, a total flux density of 0.4\,mJy and full width at half maximum (FWHM) $\sim 8$ mas.  This component is located on the south-eastern side (SE) with respect to the Gaia position.  The nearest observations with the VLBA on days 18, 23, and 27 all resulted in non-detections, which are consistent with the EVN result given the different sensitivity achieved in these runs.

The following two VLBA observations reveal emission at 5 GHz in the SE region: the flux densities are 1.8 and 2.4 mJy, on days 31 and 36 respectively, and the FWHM between 7 and 10 mas.

By day 44 (EVN at 1.6 GHz, Fig.~\ref{f.day44}), we detect a triple source in the inner 15-20 mas, surrounded by very weak diffuse emission, and a more distant component $\sim 300$ mas to the northwest. Of the inner triple source, two components are located on the SE side, while for the first time we observe emission also on the north-western side (NW) with respect to the Gaia position and in the faraway region marked as ``K''. The total cleaned flux density on day 44 is about 3.5 mJy.  On day 55 (VLBA at 1.6 GHz), we reveal a 0.6 mJy component in the north-western part of the shell, plus an  additional 1.2 mJy distributed over a wide arc-like feature in the south-eastern part.

On day 70 (EVN at 5 GHz, Fig.~\ref{f.day70}), the inner region is resolved in a structure with complex asymmetric shell-like morphology: the south-eastern front is brighter than the north-western side; the total flux density has grown to almost 5 mJy. No significant emission is present in the distant component K detected earlier at 1.6 GHz.

\begin{figure}
\includegraphics[width=\columnwidth]{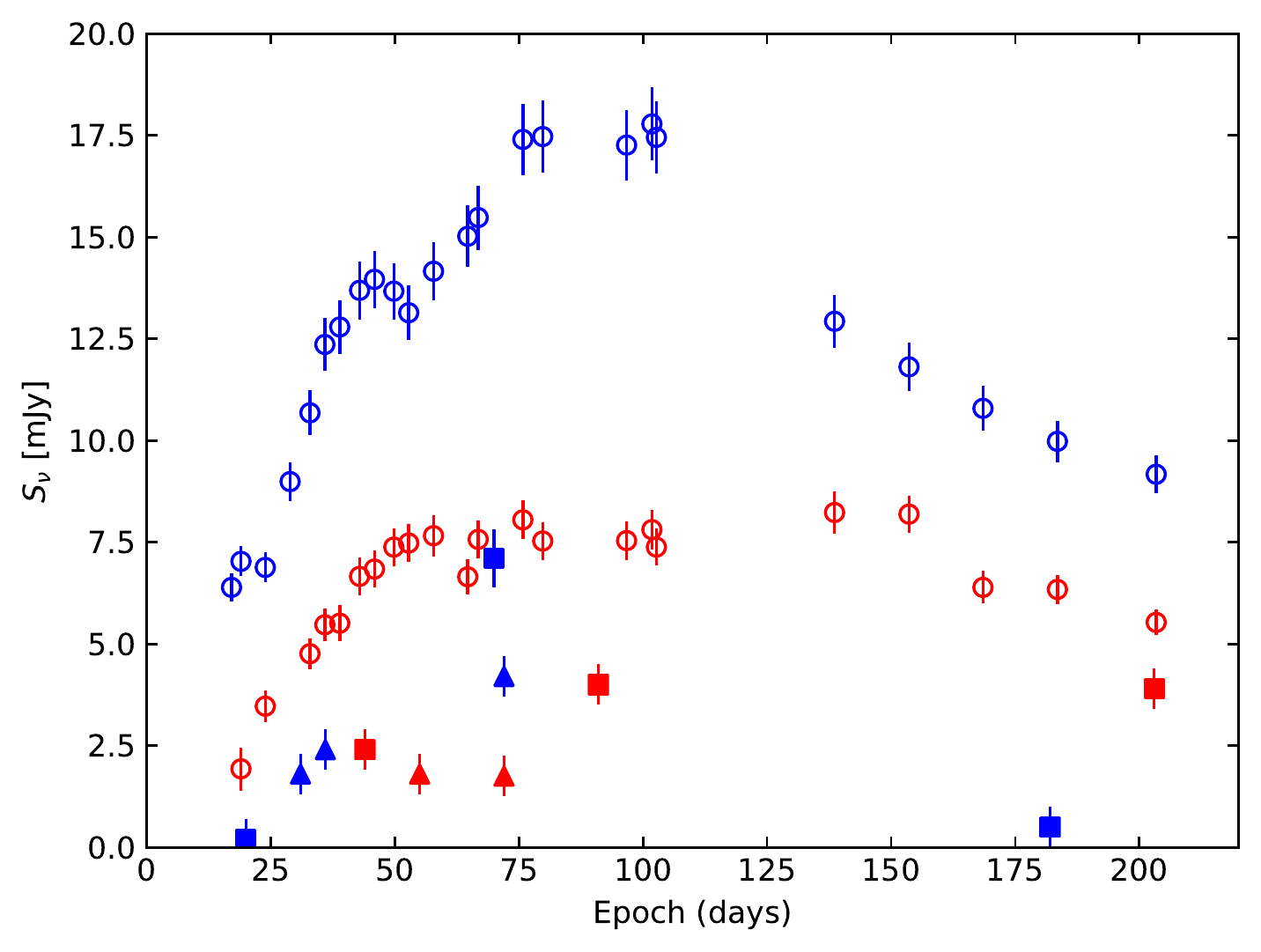}
\caption{\vv\ light curve with VLA \citep[empty circles, from][]{Chomiuk2012} and VLBI (filled symbols) data from the present paper: filled squares and triangles represent EVN and VLBA data, respectively. Red is for 1.6 GHz (1.8 GHz for the VLA data), blue for 5 GHz. \label{f.lightcurve}}
\end{figure}

On day 72, we have the first and only simultaneous detection at 1.6 and 5 GHz, obtained with the VLBA in images produced with natural weighting schemes (third row of Fig.~\ref{f.vlbi_in}).  In the 1.6 GHz data, we reveal both the north-western and the south-eastern fronts of the shell (with similar flux densities of $\sim 0.6$ mJy each); at 5 GHz, we only detect the emission from the south-eastern side, with a total flux density of 3.3 mJy.  This is the final VLBA observation; we did not detect significant emission from the faraway north-western component K in any of the VLBA datasets. 
On day 91 (EVN, 1.6 GHz, Fig.~\ref{f.day91}), we detect again all the features:\ the inner components, distributed in a shell-like structure, brightest at its south-eastern edge; the distant emission region K, with a  flux density of $\sim 0.5$ mJy, and some weak, diffuse emission.  The total flux density at this epoch is around 5 mJy.

In the final EVN epochs (day 182 at 5 GHz and day 203 at 1.6 GHz, respectively), the flux density decreases back to rather low levels, and the morphology starts to become confused with the noise again. On day 182, we barely detect 1 mJy of flux density at 5 GHz from the inner region, while on day 203 a few mJy are present at 1.6 GHz, partly in the inner region and partly in the outer component K.

In total, the flux density integrated across all the structures detected by the VLBI observations shows a clear evolution with time, with an initial rise and a late decay. In Fig.~\ref{f.lightcurve}, we show the light curve based on the total flux density in the VLBI images. The 5 GHz data (blue symbols) show a rise up to day 70, and then start declining. The 1.6 GHz measurements (red symbols) are less variable, although we note that the earliest detection at this frequency was only on day 44, which also indicates that the source must have been weaker in the early stages of development. In the same figure, we also show for comparison the light curve at the nearest frequencies, as obtained by \citet{Chomiuk2012} with the \textit{Jansky} VLA. The trends are roughly in agreement, with the 5 GHz data showing larger variations and an earlier peak than the 1.6 GHz ones. However, there is a clear difference in the total flux density, which in our clean images is significantly smaller (at least by a factor of a few) of the nearest (in time and frequency) VLA measurement (see also \ref{s.structure} and the discussion section). 

\begin{figure*}
\center \includegraphics[width=0.95\textwidth]{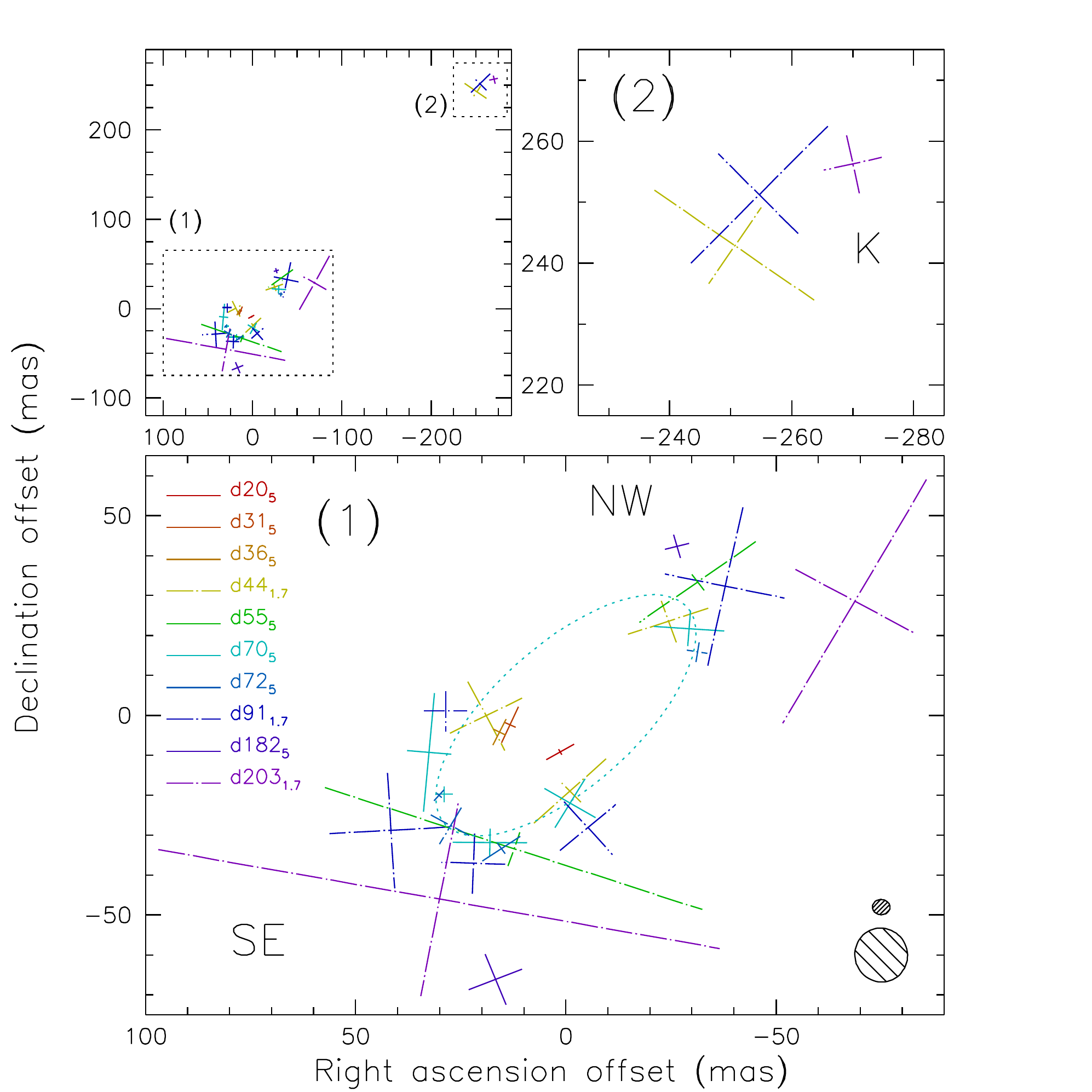}
\caption{Model-fit components. Top left: entire field; top right: zoom in the outer component region; bottom: zoom in the central shell region. Crosses indicates the position, size, and orientation of the model-fit Gaussian components; solid and dot-dash lines represent 5 GHz and 1.6 GHz observations, respectively. The different colours show different epochs as described in the the legend; the subscript indicates the observing frequency. The dotted curve shows the best-fit ellipse to the set of components at day 70. The reference position for all fields is at  R.A.\ 21h 02m 09.81700s,  Dec.\ $+45^\circ$ 46$^\prime$ 32$^{\prime\prime}$.68253. The hatched (almost circular) ellipses in the bottom right corner show the HPBW at 1.6 GHz (from day 91) and 5 GHz (from day 70), for reference. \label{f.modelfit}}
\end{figure*}

\begin{figure}
\center \includegraphics[width=\columnwidth]{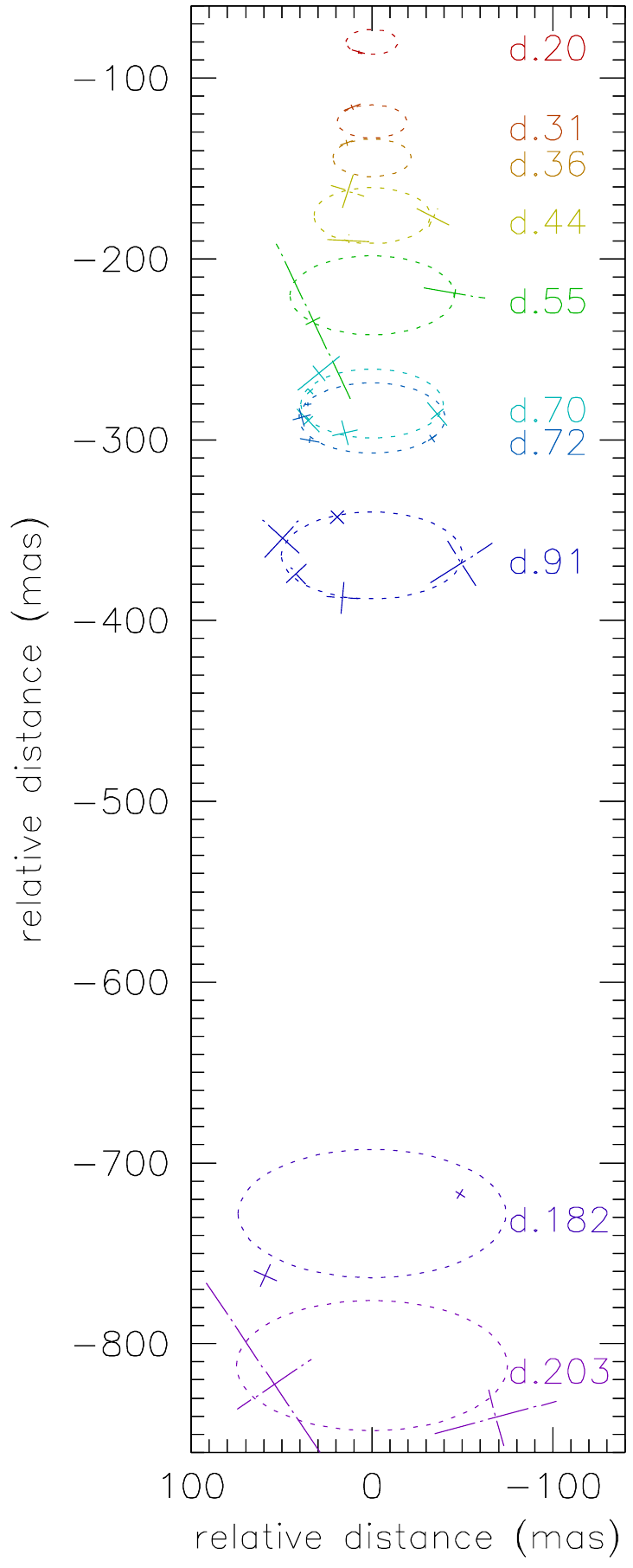} 
\caption{Ellipse fits to the sets of Gaussian components. Each ellipse is rotated clockwise by $44^\circ$ and shifted by a vertical space proportional to the time elapsed since the nova. Crosses show the sets of Gaussian components at each epoch, rotated clockwise by $44^\circ$; solid and dot-dash lines represent 5 GHz and 1.6 GHz observations, respectively. \label{f.ellipses}}
\end{figure}

\subsection{VLBI structure and model fits}\label{s.structure}

The small structures detected in this study, with angular sizes of the order of $\sim$10 mas, could not be resolved in the VLA observations like those presented by \citet{Chomiuk2012} that are characterised by a much wider beam size of several arcseconds. Conversely, our observations are not sensitive to large scale structures (100-1000$\times$ wider than our beam), which act like a uniform background completely resolved out at the angular scales sampled by the VLBI baselines.  This allows the finest structural details to emerge, as we aimed for with the EVN and VLBA.  This naturally accounts for the difference in the flux density level of the two types of observations, $\sim 5-20\times$ larger for VLA that merges all the emission within a much wider angle compared to VLBI arrays, that measure only the flux density from the smaller scale structures. Therefore, the components discussed in the following are representative only of the most compact and densest locations in the system. 

To describe  such structures with measurable quantities, we have modelled the $(u,v)$-data in Difmap with 2-d Gaussian components. In general, we tried to use elliptical components when the signal-to-noise ratio was adequate. For fainter features, we resorted to circular components. We list in Table~\ref{t.modelfit} the resulting parameters for each component:  polar coordinates $r,\theta$; flux density $S_\nu$; major and minor axis and position angle (p.a.) $a, b, \phi$; location within the source structure: we use SE to indicate components located at the south-eastern edge of the inner shell, NW for those at its north-western side, and K for the distant knot up to the north-west; these regions are also marked in the EVN images shown in Figs.~\ref{f.day44}-\ref{f.day91}.

We also show the various components in Fig.~\ref{f.modelfit}. In the top left panel, we show the entire field with crosses representing the orientation and size of each feature; we use different colours and line styles to represent the set of components at each different epoch and frequency. In the top right panel, we show a zoom on the distant knot region, where only 1.6 GHz EVN observations reveal a component. In the main bottom panel, we enlarge the region of the main shell-like structure. In total, four to six components are needed to describe the EVN observations on days 44, 70, and 91, in which the structure is most complex. Only one or two components are present in the remaining EVN data and in all the VLBA observations. 
In each epoch, the total flux density is quite evenly spread among the various components. The ratio between the flux density of any two components in the shell region is generally around unity and it never exceeds $\sim4$. 

Given the rapid evolution of the source and the different angular resolution of the array at 1.6 and 5 GHz, it is quite hard to identify each individual component across the various epochs. However, there is a good match of the whole set of model fit components between the various epochs. The sets of Gaussian model fit components of the nova shell seem to be aligned along ellipses. To quantify these structures, we have fit the Gaussian model components with a sequence of ellipses. As some epochs only have very few distinct model fit components, we fixed the centre position $(x_0,y_0)$ as well as the eccentricity $\epsilon$ of all ellipses to one common value ($x_0= 21\mbox{h}\ 02\mbox{m}\ 09.81700\mbox{s}$, $y_0= +45^\circ\ 46^\prime\ 32^{\prime\prime}.68253$, $\epsilon = 0.48$). Only the semi-major axis, and thus the expansion speed were left as free parameters.  The  direction of the major axis was chosen to align with the outer north-western component ($\theta = -44^\circ$). For illustration purposes, we show the best fit ellipse for day 70  with a dotted line in Fig.~\ref{f.modelfit}. Each ellipse corresponding to every single epoch is also shown in the set of panels presented in the Appendix (Figs.~\ref{f.vlbi_in} and \ref{f.vlbi_out}). The separation between the ellipse centre and the Gaia position is 16.7 mas in PA 118$^\circ$.

\begin{table}
   \centering
   \caption{
\label{t.shell}Fit to the size of the major axis of the shell representing the inner structure.}
\begin{tabular}{cccc}
\hline
   \hline
Time since nova & \multicolumn{3}{c}{Semi-major axis}\\ 
(days) & (mas) & ($10^{13}$ cm) & (AU) \\
\hline
20	&	$	14.3	\pm	2.5	$	&	$	58	\pm	10	$	&	$	39	\pm	7	$	\\
31	&	$	19.1	\pm	3.2	$	&	$	77	\pm	13	$	&	$	52	\pm	9	$	\\
36	&	$	21.7	\pm	3.0	$	&	$	88	\pm	12	$	&	$	59	\pm	8	$	\\
44	&	$	32.1	\pm	1.4	$	&	$	130	\pm	6	$	&	$	87	\pm	4	$	\\
55	&	$	45.5	\pm	8.6	$	&	$	184	\pm	35	$	&	$	123	\pm	23	$	\\
70	&	$	39.5	\pm	3.0	$	&	$	160	\pm	12	$	&	$	107	\pm	8	$	\\
72	&	$	40.4	\pm	3.6	$	&	$	163	\pm	15	$	&	$	109	\pm	10	$	\\
91	&	$	50.1	\pm	2.3	$	&	$	202	\pm	9	$	&	$	135	\pm	6	$	\\
182	&	$	74.0	\pm	4.9	$	&	$	299	\pm	20	$	&	$	200	\pm	13	$	\\
203	&	$	75	\pm	30	$	&	$	303	\pm	121	$	&	$	203	\pm	81	$	\\
\hline
\end{tabular}
\end{table}

\begin{table}
   \centering
   \caption{
\label{t.outer}Fit to the position of the outer component.}
\begin{tabular}{cccc}
\hline
\hline
Time since nova & \multicolumn{3}{c}{Radius}\\ 
(days) & (mas) & ($10^{13}$ cm) & (AU) \\
\hline
44	&	$	349	\pm	3	$	&	$	1410	\pm	12	$	&	$	942	\pm	8	$	\\
91	&	$	358	\pm	4	$	&	$	1446	\pm	16	$	&	$	967	\pm	11	$	\\
203	&	$	372	\pm	9	$	&	$	1503	\pm	36	$	&	$	1004	\pm	24	$	\\
\hline
\end{tabular}
\end{table}

\subsection{Evolution with time: component motions}
\subsubsection{Inner structure}\label{s.time-in}

We show the results of the fits as a function of time in Table \ref{t.shell} and Fig.~\ref{f.ellipses}. For epochs with more than two components, the uncertainty on the ellipse semi-major axis $r$ is directly provided by a least squares fit. For epochs with only one (or two) components, the uncertainty is given by the (mean) size of the Gaussian component(s) along the line between the component centre and $(x_0,y_0)$.

Both the visual (Figs.~\ref{f.modelfit} and~\ref{f.ellipses}) and quantitative (Table \ref{t.shell}) inspection of the model fit component positions reveals a clear overall expansion of the source inner shell-like structure. The ellipse semi-major axis grows from $(14.3\pm2.5)$ mas, or $(5.8 \pm 1.0)\times 10^{14}$ cm, at day 20, to $\sim 75$ mas, or $\sim 3\times10^{15}$ cm, at day 203. We also reveal a proper motion for the outer component, whose distance from our reference point is increasing as a function of time; the results for this component are presented in Table \ref{t.outer} and Sect.~\ref{s.time-out}.

The densest time coverage and the most well constrained fits are in the time range between day 20 and 91. In this interval, a weighted least square linear fit of the ellipse semi-major axis vs.\ time based on the values reported in Table~\ref{t.shell} provides a speed of $v_{20-91}=(2\,100 \pm 300)$\,\kms. In addition, a simple estimation of the mean  velocity in the first 20 days  based on the size of the ellipse at the first epoch and assuming constant velocity provides $v_{0-20}=(3\,300 \pm 600)$\,\kms. At the latest epochs (day 182 and 203), the signal to noise ratio becomes quite small again, both because of the expansion and dimming of the source and the somewhat lower data quality. Accordingly, the constraints on the velocity become less tight but still consistent with a continued expansion, albeit with slower velocity, i.e.\ down to an average of $v_{91-182}=(1\,200 \pm 100)$\,\kms\ between days 91 and 182, and as small as $\sim 200$ \kms\ if we consider only the last two epochs. 

\begin{figure}
\includegraphics[width=\columnwidth]{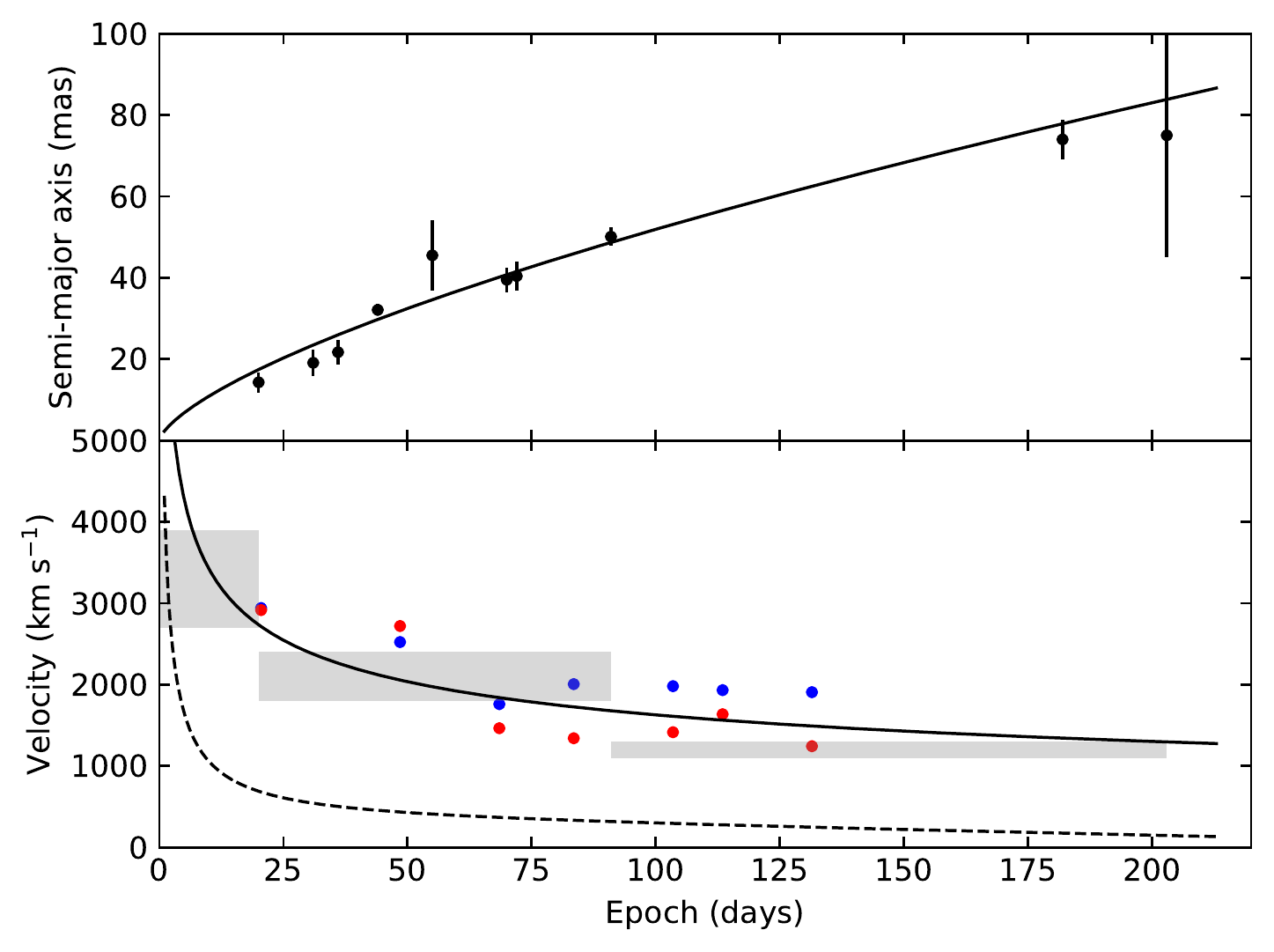}
\caption{Top panel: semi-major axis of the ellipse as a function of time; the solid line indicates the best-fit power-law relation $r \sim t^p$, $p=0.68\pm0.04$. Bottom panel: ejecta advance velocity obtained from the expansion fit (solid line) and from simple averages in separate time intervals (grey shaded rectangles); for comparison, we plot also the H$\alpha$ FWHM time evolution by \citet[][dashed line]{Munari2011}; the blue (red) points indicate the maximum positive (absolute negative) radial velocity for H$\alpha$ from \citep{Shore2011}. \label{f.expansion}}
\end{figure}

\begin{table*}
   \centering
   \caption{
\label{t.spectrum}Shell spectral index.}
\begin{tabular}{lcccccccc}
\hline
\hline
 & \multicolumn{3}{c}{VLBA} & \multicolumn{5}{c}{EVN} \\
Component & $S_{\rm 1.6,\ d72}$ & $S_{\rm 5,\ d72}$ & $\alpha$ & $S_{\rm 1.6,\ d20}$ & $S_{\rm 1.6,\ d91}$ & $\langle S_{\rm 1.6}\rangle$ & $S_{\rm 5,\ d44}$  & $\alpha$ \\
 & (mJy) & (mJy) & & (mJy) & (mJy) & (mJy) & & \\
\hline
Total	& 1.75 & 4.7 &  0.9 & 2.4 & 4.6 & 3.5 & 8.3 & 0.8 \\
SE   & 1.0 & 4.2 & 1.3 & 1.5 & 3.0 & 2.25 & 7.1 & 1.0 \\
NW & 0.75 & $<0.5$ & $<-0.3$ & 0.9 & 1.6 & 1.25 & 1.2 & $-0.1$ \\
\hline
\end{tabular}
\end{table*}

In Fig.~\ref{f.expansion} (top panel), we show the trend of the ellipse semi-major axis $r$ as a function of time $t$, overlaid with a weighted least square power-law fit. We use all data over the interval between day 0 and day 203, with the expression $r\propto t^p$, and obtain $p=0.68\pm0.04$. Taking the derivative of the size, and considering our assumed distance of $d=2.7$ kpc, we get the associated advance velocity.  We plot this velocity in the bottom panel, overlaid to shaded boxes which indicate the simple mean values for the three intervals $t\le20\,$d, $20\,\mathrm{d}\le t \le 91\,\mathrm{d}$, $t \ge 91\,$d.  The two methods are mostly consistent with each other: the best-fit power-law trend does a better job to describe a continuous evolution, instead of the unrealistic jumps based on the constant velocity assumption; on the other hand, this fit  diverges for $t\to0$ (as $p<1$), meaning that the starting velocity can not be well constrained. 

In the bottom panel of Fig.~\ref{f.expansion}, we also show the  evolution of the spectroscopic (hence, distance independent) measurements from \citet{Munari2011} and \citet{Shore2011}. We show in particular the fit to the FWHM (in \kms) of the broad component of H$\alpha$  \citep[][]{Munari2011}, with the dashed line, and the maximum H$\alpha$ radial velocities reported by \citet[][]{Shore2011}, with the coloured dots. Not shown in our figure are the He~\textsc{II} maximum radial velocities \citep{Shore2011}, which show a power-law time dependence with an exponent of $-0.84 \pm 0.05$ on the approaching side and of $-0.31\pm0.05$ and the receding side. The latter exponent is consistent, within the uncertainty, with the index of the derivative of our expansion power law ($p-1=-0.32\pm0.04$) in a similar time range. While the expansion velocity is distance dependent, the power-law is not.

We also consider an alternative scenario in which the reference position for the motion of components is located at the coordinates determined by extrapolating the Gaia results.  In this scenario, the SE and NW regions are still advancing on opposite sides of each other, but NW would be significantly closer than SE to the origin.  As a result, the motion of the ejecta would be quite different on the two fronts, with velocities on SE about a factor $2\times$ higher than the other side.  The results obtained under this scenario are less informative, both because it is less well constrained (the motions on each of the two sides are fit separately) and because the Gaia position, representative of the Mira rather than the WD, is not necessarily the best candidate to represent the origin of the motion of the ejecta.  Therefore, we only present it in the Appendix (Sect.~\ref{s.appendixMotion}) to provide a term of comparison for our primary scenario based on the growing ellipses.

\subsubsection{Component ``K''}\label{s.time-out}
In Table \ref{t.outer}, we further report the results of the fit to the motion of the outer component, which is consistent with rather low projected speeds ($0.15\pm0.06$ mas day$^{-1}$, or $700\pm280$\,\kms) in the observed time range. Given the observed angular speed and separation, we can estimate a distance independent epoch of ejection about 6.7 years before the 2010 nova, i.e.\ in late 2003, or somewhat later for a decelerating motion. 

Since this component is far ($\sim 940$ AU) from the nova, if we assumed that the component were ejected at day $\sim 0$ from the shell origin, we would derive a quite large average speed for the initial period, $v\sim30\,000\, \mbox{\kms} \sim 0.1c$ (projected). Since a constant speed in this period is unlikely, the component would have to have been ejected with relativistic speeds. Combined with the inspection of the optical light curves (see Sect.~\ref{s.discussion-3}), an origin in a previous phase of activity as described in the previous paragraph seems a much more solid scenario.

\subsection{Spectral properties}

The dual frequency observations at our disposal allow us to discuss also the spectral properties of the resolved emission, although with a few caveats: the EVN observations at 1.6 and 5 GHz are not simultaneous, which is an issue given the structural and flux density variability; the VLBA multi-frequency observations, on the other hand, have simultaneous data, yet we have only one detection at the two frequencies in the same epoch; moreover, as typical for VLBI arrays, it is not possible to obtain matched $(u,v)$-coverages anyway, which introduces further uncertainty on any quantitative estimate of a spectral index.

With all the above caveats in mind, we report in Table \ref{t.spectrum} both the simultaneous VLBA spectral index obtained at day 72, and the average EVN spectral index obtained by combining the mean of the 1.6 GHz observations on day 44 and day 91 and the intermediate 5 GHz observation on day 70. Despite the significant differences in time and angular resolution, the results are in good agreement. The overall spectral index of the inner features is rising ($\alpha\sim0.8$) but the two opposite fronts are characterised by different properties. The south-eastern front spectrum has a clear rising trend ($\alpha\sim1.1$). Since this side of the shell is brighter, this accounts for the overall spectrum of the entire shell. On the other hand, the north-western component has a much flatter (if not steep) spectrum. 

The spectral index asymmetry shown in Table \ref{t.spectrum} for the time range $44 \mathrm{d} \le t \le 91 \mathrm{d}$ is also supported qualitatively by the earlier 5 GHz VLBA detections (day 31 and day 36), both localised in the south-eastern side. We note that the finding of a rising spectrum is at least qualitatively robust against the use of non-matched $(u,v)$-plane coverage, since the bias is in general toward an overestimation of the flux density at low frequency, due to the sampling of small spatial frequencies (i.e.\ larger size structures).

As for the faraway isolated component, it is only detected with the EVN at 1.6 GHz. While at first sight this would suggest a steep spectrum, we note that the component is extended, with a major axis as large as $\sim 30$ mas.  Given the smaller beam size and higher noise level for the 5 GHz observations, it is possible that the emission is resolved out at 5 GHz even for positive values of the spectral index. As a matter of fact, we can only put a very loose constraint on the spectral index of $\alpha \le +2.9$.

\section{Discussion\label{s.4}}

This section is organised as follows:  we first review the literature to set the background about the linear scales involved for this system (Sect.~\ref{s.discussion-1}). We then devote Section~\ref{s.discussion-2} to the geometry, kinematics, and spectral properties of the VLBI components in the inner structure, proposing a 3-d geometrical interpretation and a discussion of the emission mechanisms and the associated physical conditions. Finally, the properties and possible origin of component K are the subject of Sect.~\ref{s.discussion-3}.

\subsection{Linear scales}\label{s.discussion-1}

The linear scales involved in this system are determined by the distance $d$ of the binary from us and the separation $a$ between the white dwarf and the Mira.  As outlined in the introduction, there is some tension between estimates of $d$ based on different methods. The period-luminosity relation for Miras provide a distance in the range between 1.7 and 2.7 kpc \citep{Munari1990,Kolotilov2003}. The presence of Na~\textsc{I}~D absorption in the optical spectra, likely caused by the Perseus spiral arm, suggests a larger distance, $d\ge3.0$ kpc \citep{Chomiuk2012}.  Our adopted choice of 2.7 kpc is a reasonable compromise, which however implies a systematic uncertainty on all the linear scales and projected velocities of order $\sim20\%$.

In terms of orbital size $a$, \citet{Abdo2010} estimated a separation between the white dwarf and the red giant of about $a=10^{14}$ cm, i.e.\ $a=6.7$ AU. This estimate is based on the delay between the optical and gamma-ray emission ($3-4$ days) and the velocity derived from the optical lines ($3200\pm345$ \kms). However, our results show that this is the average velocity during the first 20 days, so it is likely that the initial velocity is larger, as also suggested by the power-law fit to the expansion.  It is thus likely that the real separation is wider, which would be in better agreement with findings based on the study of optical periodicity:  e.g., \citet{Munari1990}, from a period of 43 yr, suggest  17 AU for a $1M_\odot-1.5M_\odot$ binary \citep{Chomiuk2012} or even 20-25 AU if the red giant has a mass of $4-8 M_\odot$ \citep{Nelson2012}. The long term radial velocity monitoring of the Mira in \vv\ performed in the near-IR by \citet{Hinkle2013} has failed to reveal any orbital motion superimposed to the regular pulsation, a fact interpreted by the authors as indicative of an orbital period $\geq$100 years, that for $M_1 + M_2 = 3 M_\odot$, corresponds to $a\geq$31 AU (=11.5 mas).

The projected separation between the Gaia position and the assumed centre of expansion is consistent with the upper values of this range. As detailed in Appendix~\ref{s.appendixGaia}, the statistical uncertainty of the extrapolated Gaia position is rather small (1.4 mas) at the epoch of the nova explosion and the extrapolation further back in time (to the VLA detection in 1993) suggests that also systematic uncertainties should be small. It is less easy to estimate an uncertainty on the position of the centre of expansion and therefore on the separation. The components on the SE side are in general closer to the centre of the ellipse than those on the NW front at the same epoch, so the $\sim16$ mas value should be taken as an upper limit to the projected separation.

\subsection{Properties of the inner structure}\label{s.discussion-2}

Most reports of radio emission from \vv\ after the 2010 nova event show it is unresolved on angular scales larger than about one arcsecond \citep{Chomiuk2012}. 
Our images clearly resolve the radio emission in \vv\ after the nova event and show it is distributed in at least two main structures:\ the distant isolated feature (K), likely related to a previous episode of activity (see Sect.~\ref{s.discussion-3}), and the elongated shell-like set of inner components (SE and NW), clearly associated to the 2010 nova event, and discussed in this subsection.  In any case, these two main structures are not unrelated, as indicated by the alignment between the direction of expansion of the inner components and the p.a.\ of the outer feature.  This suggests a preferred axis for the whole system, which is naturally interpreted as the polar axis of the binary orbital plane. The two fronts described by the SE and NW sets of components show how the ejecta advanced preferentially along this polar axis (in opposite directions), rather than at larger zenithal angles (nearer to the orbital plane), where the density of the Mira wind may be higher and more difficult to penetrate \citep{Walder2008,Mohamed2012,Booth2016}.

This first finding provides a direct proof that the ejecta are expanding aspherically, as earlier indicated by several arguments: \citet{Shore2011} present high-resolution optical spectra with asymmetric structure in several emission lines, with larger velocities in red wings than in blue ones and different slopes of the power-law time evolution on the opposite sides; \citet{Nelson2012} also developed an asymmetric model on the basis of the presence of multiple degrees of ionisation equilibrium in the X-ray emission; \citet{Chomiuk2012} used this model to interpret the thermal emission from the ionised envelope of the Mira.  If the two fronts of the emission are advancing along the polar axis, the derived velocities of $\sim 2\,000$ \kms\ at days $30-90$ imply that we are seeing the system nearly edge-on; otherwise, the de-projected velocities would become inconsistent with those derived for the blast wave from the Suzaku X-ray spectrum \citep{Nelson2012}, which indicates velocities between 1\,450 and 1\,800 \kms.  The strong asymmetry of the line width profiles mentioned earlier was interpreted by \citet{Shore2012} with a geometrical model in which the WD is seen close to superior conjunction, through a substantial portion of the Mira wind.  Our images provide a nice direct confirmation of such scenario. If the WD-Mira separation has also a significant perpendicular component, in addition to the projected one which is already of several AUs, this means that the orbital radius value is likely in the upper range of those proposed in the literature.

With a striking analogy with RS Oph \citep{O'Brien2006,Rupen2008}, our observations further reveal a morphological asymmetry between the two fronts of the expansion.  On one side (East in RS Oph, SE in our case, see in particular the EVN image on day 70 in Fig.~\ref{f.day70}), the structure is shaped as a ring associated with the shock wave generated by the outburst \citep{Rupen2008}. The analogy also extends to the opposite side of the ring, which is fainter for both sources, and in the presence of a more distant component (on the same side as the inner bright shell in RS Oph, on the opposite side in \vv).  However, the similarity breaks  when the spectral properties are considered.  The shell in RS Oph has a steep spectrum (flux density decreasing as a function of frequency), which is straightforward to interpret as synchrotron emission from the shocked ejecta, also considering the high brightness temperature $T_\mathrm{b}$.  In \vv, we find flat-rising radio spectra, which are much harder to interpret.    
The brightness temperatures observed for our components are typically of the order of a few $\times 10^5$\,K, with one case in which $T_\mathrm{b}$ gets as high as several $\times10^6$\,K (the 4.0-mas, 1.6-mJy 5-GHz component in the south-eastern front at day 70). While not as high as in RS Oph, this $T_\mathrm{b}$ is unlikely to be due to thermal emission from hot gas.  Although \citet{Nelson2012} showed that hot, shocked, X-ray emitting gas was
indeed present in V407 Cyg at the epoch of our radio observations, the X-ray emission measure of that plasma indicates that there was nowhere near enough of it to generate the radio emission we observed.    For example, for the 1.6 mJy of 5-GHz radio emission from the 4.0-mas knot on day 70 to have been thermal emission from the X-ray emitting gas, it would have needed a density of around $10^8 \mathrm{cm}^{-3}$.  But such a knot of high-density, $3 \times 10^7$ K plasma would have produced orders of magnitude more X-ray emission than was observed. Therefore, in the lack of meaningful polarisation information, we can only speculate that the radio emission is non-thermal, yet with a spectral index far from the canonical one.  Synchrotron emission with such unconventional spectral index is also what has been reported in at least two classical novae, i.e.\ V5589 Sgr \citep{Weston2016} and V1324 Sco \citep{Finzell2018}, with the latter also detected in gamma rays \citep{Cheung2012,Ackermann2014}. 

While the \textit{asphericity} is naturally intrinsic to the system, it is not clear whether this ejecta \textit{asymmetry} in brightness and spectrum is intrinsic, driven by the different interaction with the circumstellar medium, or caused by different absorption along the line of sight. The fact that the SE front is brighter and has a positive $\alpha$ could be an indication that these ejecta are moving into higher density material towards the Mira, so they become denser and brighter but also more absorbed at low frequency. However, the coordinates obtained by Gaia seem to indicate that the Mira is located on the NW side with respect to the ellipse centre.  Even in the secondary scenario in which the expansion is centred on the Gaia position and asymmetric (Appendix~\ref{s.appendixMotion}), the highest velocity on the SE side would indicate that to be the less dense side.  Furthermore, these simple schemes could  be complicated by the density waves imparted by pulsation to the steady wind outflow from the Mira: at 10 km sec$^{-1}$ outflow speed and 745 day pulsation period, the waves are separated by $\sim$4 AU, smaller than the orbital separation in \vv. 
In the end, it seems reasonable to conclude that the polar axis lies almost on the plane of the sky, and the asymmetries on the two fronts are likely intrinsic to inhomogeneity in the explosion.

One could consider an alternative scenario in which the VLBI features are advancing along the equatorial plane, with the WD-Mira vector either in the plane of the sky (in disagreement with the asymmetries in the optical line profiles) or orthogonal to it (and with the WD behind the Mira).  Our velocities at late time are however significantly larger than those seen in the optical spectra, which is inconsistent with the advance against higher density gas in the equatorial plane. This, combined with the analogy with RS Oph and the presence of the outer component K in the same p.a.\ (see next subsection), lends support to the previously described scenario in which the explosion resulted in a jet/collimated outflow along the polar axis.

\subsection{On the nature of component K.}\label{s.discussion-3}

The isolated component K is well aligned (p.a.\ $\sim -45^\circ$) with the axis of the bipolar expansion of the inner lobes, like a polar jet.  The geometrical arrangement reminds of that adorning the symbiotic binary He 2-104, the {\it Southern Crab}, in the spectacular HST images presented by \citep[][their Figure~1]{Corradi2001}: two bipolar lobes expanding perpendicularly to the equatorial plane and, at a greater distance along their main axis, two very compact polar jets.  Spatio-kinematical modelling of the lobes and jets in Hen 2-104 indicated a similar age for them ($\sim$5700 yrs ago), even if not necessarily an origin in one and the same eruption.  As for \vv, Hen 2-104 harbours a Mira of long pulsation period, orbited at great distance by a WD companion.  The great orbital separation is inferred by the fact that (i) the Roche lobe around the Mira contains enough dust to completely block the visibility of the Mira itself at optical
wavelengths, while the WD is seen free from such dust, and (2) the dust around the Mira is sufficiently distant from the WD to survive the hostile hard radiation field created by the stable hydrogen-burning going on at the surface of the WD since earliest available photographic plates.

As noted above, the $\sim$700 \kms\ projected velocity and the $\sim$940 AU projected separation suggest that component K left the central binary $\sim$7 years before our VLBI observations, or $\sim$2003.  At that time \vv\ was going through an active phase powered by accretion onto the WD. Normally very faint (pulsation-averaged $B\geq19$ mag; \citealt{Munari1990}), in 1993 \vv\ begun slowly but steadily to rise in brightness, with its spectrum turning gradually from a normal field Mira to that of a symbiotic star with strong Balmer, He~\textsc{I} and [O~\textsc{III}] emission lines, as first noted by \citet{Munari1994}.  It kept increasing in brightness and excitation level, reaching a peak $B\sim13$ mag in 1999 and than slowly declining to $B\sim14.5$ mag in 2003 and $B\sim16$ mag shortly before the onset of the nova eruption in March 2010.  The polar K component has therefore left \vv\ during this accretion episode, slightly after its peak brightness.  The only other known active phase displayed by \vv\ was in 1936, but that one was shorter ($\sim$2 years) and fainter (peak $B\sim14.5$ mag). It is worth noticing that the accretion phase started in 1993 turned \vv\ from a radio-quiet into a radio-loud source. In fact, VLA observations in 1989 by \citet{Seaquist1993} posed an upper limit of 0.06 mJy at 8.4 GHz to emission from \vv, while VLA observations for May 14, 1993 (as reprocessed by \citealt{Chomiuk2012}, see also Appendix~\ref{s.appendixGaia}) detected emission at 1.2 mJy again at 8.4 GHz (therefore an increase of at least a factor of 20).

It is not unusual for jets to leave symbiotic binaries during active phases.  Signatures of jets have been visible in high resolution spectra \citep[eg.\ Z And, Hen 3-1341][]{Burmeister2007,Tomov2000} or in high resolution imaging \citep[eg.\  R Aqr,][]{Schmid2017}, up to the spectacular $-$6000 \kms\ case of MWC 560 when it erupted in 1990 \citep{Tomov1990}.
The case most pertinent to \vv\ seems however that of CH Cyg.  This is an
eclipsing symbiotic binary of very long orbital period (15 years),
containing a semi-regular pulsating AGB of the same $\sim$M7III spectral type as for \vv.  CH Cyg has been in quiescence for decades, until in the
1960's when it began to awake rising in brightness and displaying ever
increasing emission lines caused by increased accretion onto the WD
companion to the red giant.  Around the time of peak brightness, in 1984 CH Cyg emitted a jet first observed in the radio \citet{Taylor1986} and later
resolved also at other wavelengths \citep{Solf1987}.  The observations suggested
an expansion velocity of the jet of $\sim$800 \kms, a high collimation of
the flow and an inclination axis of the jet nearly perpendicular to the line
of sight and to the orbital plane where lies the accretion disk given the
eclipsing geometry.

As a final remark, it is worth noticing that also the previous accretion episode, that of 1936, may have left a trace visible during the outburst of 2010. \citet{Shore2011}, while examining 2010 high resolution spectra of \vv\ around the Na~\textsc{I} doublet, have noticed the appearance and disappearance of feeble absorption and emission components compatible in time-sequence and velocity with an origin in circumstellar material that may have been ejected by the central binary at the time of the 1936 active state.
\citet{Shore2012} present evidence for a spectral component, in the $\left[ \mathrm{Ar} \textsc{III} \right] 7135.79 \AA$ line, associated with a low density region and an expansion velocity of about 70 \kms, which could have been produced by the eruption in the 1930's.

\section{Conclusions\label{s.5}}

A remarkable symbiotic star in itself until March 2010, \vv\ became the first nova detected in gamma rays, and it remained the only symbiotic nova revealed at $E>100$ MeV for almost a decade (\citealt{Buson2019}; see also \citealt{Franckowiak2018}). Besides the \textit{Fermi}-LAT discovery, the campaigns based on optical spectroscopy \citep{Munari2011,Shore2011,Shore2012}, X-ray observations \citep{Shore2011,Nelson2012}, and radio continuum \citep{Chomiuk2012} and spectral line \citep{Deguchi2011} observations had provided a characterisation of the initial conditions in the ejecta and in the circumstellar material. 

The high angular resolution observations presented here show for the first time a direct imaging of the inner ejecta.  These ejecta are much fainter and more compact than the extended emission from the partly ionised circumstellar gas detected by the VLA; whereas the emission from the circumstellar gas is thermal, the shocked ejecta emit via synchrotron process, with a spectral index that, although unexpected, has been reported also in classical novae \citep{Chomiuk2014,Weston2016,Finzell2018}.  The ejecta advance in opposite directions with respect to a reference position which is located within 45 AU from the Mira position, as estimated on the basis of the Gaia coordinates and proper motions.  The initial velocity ($t<20$ days) is in excess of 3\,500 \kms\ and they later slow to velocities of about 2\,000 \kms\ (between days 20 and 90).  On the south-eastern side, they are brighter and characterised by a growing radio spectrum $\alpha=+1.1$, while on the other direction they are fainter and with a flatter spectrum.  These two sets of features are interpreted as the opposite fronts of a shocked outflow that is focused along the polar axis of the binary equatorial plane; the equatorial plane is seen edge-on, with our light-of-sight nearly aligned with the direction from the Mira to the WD \citep[as in][]{Shore2011}.

We also detect a more distant component ($r=350-370$ mas, or 945-1000 AU) on the north-western side at 1.6 GHz.  This component is moving with a slower velocity of $\sim700$ \kms, and is likely related to an earlier episode of activity from the WD, in the early 2000's.  The similarities with other novae suggest that the production of long-lasting jets during active phases is a relatively common process in such systems.

\begin{acknowledgements}

We thank the JIVE Support Scientist Zsolt Paragi and the EVN scheduler Richard Porcas for their invaluable help in making the EVN observations possible. We also thank  Prof.\ Roberto Fanti, Dr.\ Domitilla De Martino, and Dr.\ Marina Orio  for useful discussion, and Dr.\ Matteo Bonato for help in producing the images. \\

The European VLBI Network is a joint facility of European, Chinese, South African and other radio astronomy institutes funded by their national research councils. The National Radio Astronomy Observatory is a facility of the National Science Foundation operated under cooperative agreement by Associated Universities, Inc.  This work made use of the Swinburne University of Technology software correlator, developed as part of the Australian Major National Research Facilities Programme and operated under licence. The research leading to these results has received funding from the European Commission Seventh Framework Programme (FP/2007-2013) under grant agreement No 283393 (RadioNet3). MG and UM acknowledge support through grant PRIN-INAF-2016. SC acknowledges the financial support from the UnivEarthS Labex program of Sorbonne Paris Cité (ANR-10-LABX-0023 and ANR-11-IDEX-0005-02). FS and KS were supported for this research through a stipend from the International Max-Planck Research School (IMPRS) for Astronomy and Astrophysics at the Universities of Bonn and Cologne. KS is partly supported by the RFBR grant 13-02-00664. Work by CCC at NRL is supported in part by NASA DPR S-15633-Y and 10-FERMI10-C4-0060. 
\end{acknowledgements}

\clearpage
\newpage

\begin{appendix}

\section{Extrapolation of the Gaia data}\label{s.appendixGaia}

\begin{figure*}
\sidecaption
  \includegraphics[width=12cm]{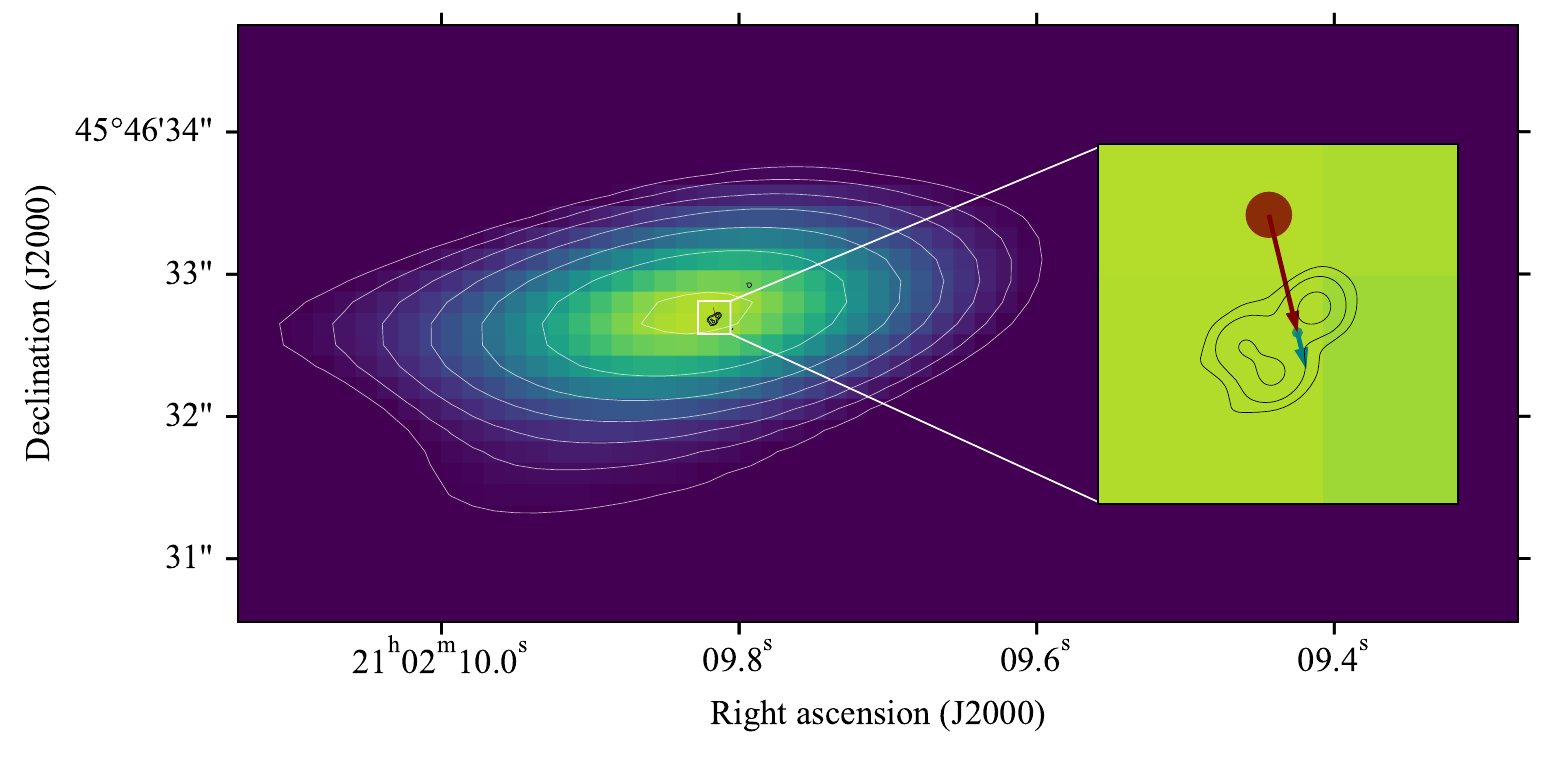}
     \caption{Archival VLA image of \vv\ at 8.4 GHz on 1993 May 18. White contours start at 0.10 mJy beam$^{-1}$ (about $3\times$ the r.m.s.\ noise level of 42 $\mu$Jy beam$^{-1}$) and increase by steps of $\sqrt 2$. The beam size is $2.6\arcsec \times 1.0\arcsec$ in p.a.\ $-82^\circ$; the peak brightness is 1.08 mJy beam$^{-1}$. 
     The inset shows the proper motion determined from the Gaia data between 1993.5 and 2010.3 (maroon arrow) and between 2010.3 and 2015.5 (teal arrow); the (almost circular) ellipses show the uncertainty on the position at the two epochs in corresponding colours, magnified by a factor $5\times$ for readability.  The black contours represent 1.6 GHz emission as observed with the EVN on day 44 (same as in Fig.~\ref{f.day44}.}
     \label{f.vla}
\end{figure*}

The second Gaia data release, Gaia DR2, contains celestial positions and the apparent brightness in $G$ for approximately 1.7 billion sources \citep{Gaia2018}. For 1.3 billion of those sources, parallaxes and proper motions are also available. These data were obtained between 2014 July 25 and 2016 May 23.  For \vv, the coordinates and proper motion at the reference epoch $t=2015.5$ are given in Table~\ref{t.gaia}. Extrapolating back to the epoch of the nova and combining in quadrature the error on the Gaia position and proper motions, we obtain the following coordinates for the epoch of the nova: (RA, Dec) = ($21^\mathrm{h}\ 02^\mathrm{m}\ 09.8156^\mathrm{s}, +45^\circ\ 46\arcmin\ 32\arcsec.6904$); ($\Delta$RA, $\Delta$Dec) = (1.4 mas, 1.4 mas); this is the position that is indicated by the cross in Figures \ref{f.day44}, \ref{f.day70}, and \ref{f.day91}.  Extrapolating the position further back to the VLA observations of 1993 presented by \citet{Chomiuk2012}, the position shifts further to the north-east and falls nicely in agreement with the peak of the radio emission, as shown in Fig.~\ref{f.vla}. The Gaia position was obtained at a time when the WD was in quiescence and therefore is representative of the Mira position rather than the mass center of the binary.  In any case, the separation between the WD and the Mira in this image would be tiny for any reasonable value.

\begin{table}
   \centering
   \caption{
\label{t.gaia}Gaia DR2 data products for \vv.}
\begin{tabular}{ll}
\hline
\hline
Quantity & Value \\ 
\hline
RA (deg) & 315.54089601822 \\
Dec (deg) & +45.77574082898 \\
$\Delta$RA (mas) & 0.1339 \\
$\Delta$Dec (mas) & 0.1526 \\
$\mu_\mathrm{RA}$ (mas yr$^{-1}$) & $-1.556$ \\
$\mu_\mathrm{Dec}$ (mas yr$^{-1}$) & $-4.474$ \\
$\Delta\mu_\mathrm{RA}$ (mas yr$^{-1}$) & 0.268 \\
$\Delta\mu_\mathrm{Dec}$ (mas yr$^{-1}$) & 0.270 \\
\hline
\end{tabular}
\end{table}

\section{VLBI images}\label{s.appendixVLBI}

\begin{figure*}
\center 
\includegraphics[width=\textwidth]{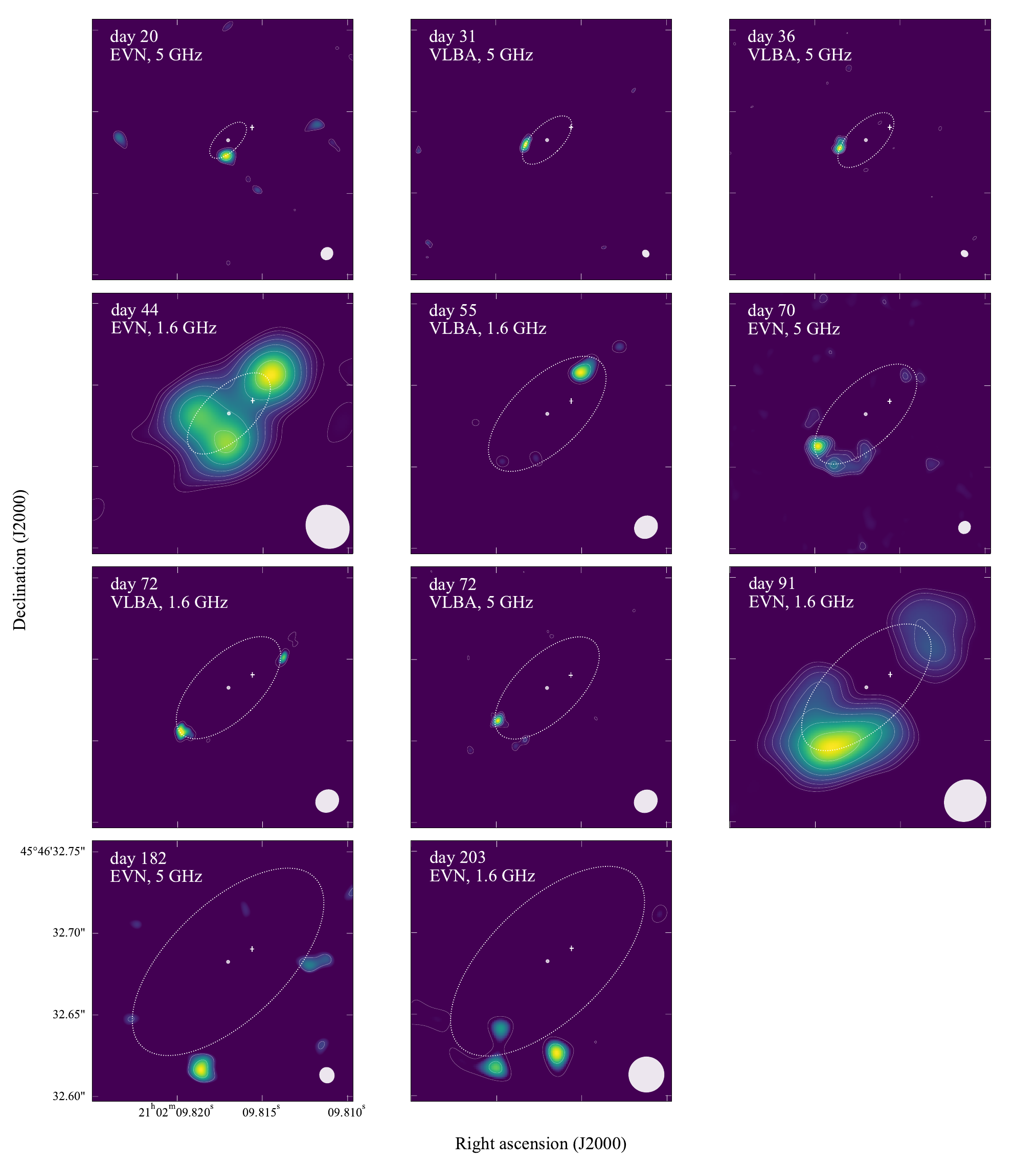}
\caption{VLBI images of the inner 160 mas $\times$ 160 mas around the center of the expansion ellipse; the ellipse and its centre are marked by the dotted line, and the white central dot, respectively.  The position of \vv\ in Gaia DR2, brought back to $t=0$ by application of proper motions, is also shown with a white cross (whose size represents the associated statistical uncertainty); the white dot indicates the position of the assumed centre of expansion. Beam sizes are shown by the filled ellipses on the lower-right corners; noise levels are given in Tables \ref{t.logevn} (EVN data) and \ref{t.logvlba} (VLBA data); the colour scale ranges from the $3\sigma$ noise level to each panel's maximum intensity.
\label{f.vlbi_in}}
\end{figure*}

\begin{figure*}
\center
\includegraphics[width=\textwidth]{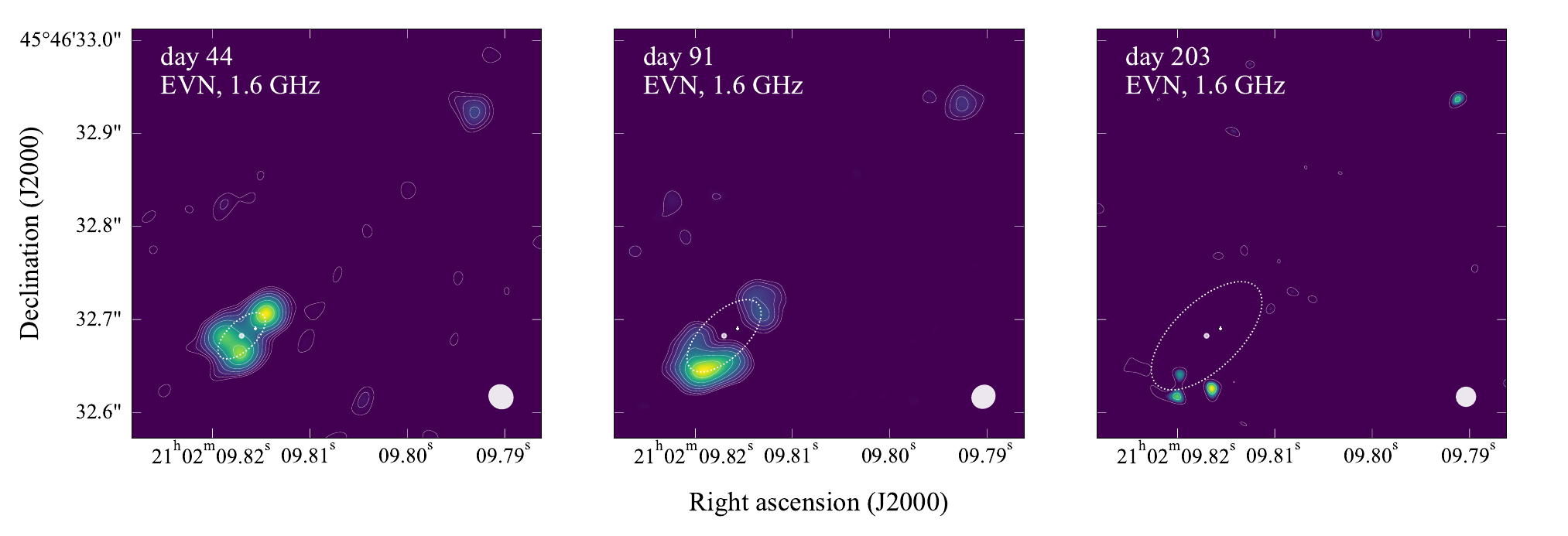}
\caption{VLBI images of the 420 mas $\times$ 420 mas field around the center of the expansion ellipse;  symbols are plotted with the same meaning as in Fig.~\ref{f.vlbi_in}. \label{f.vlbi_out}
}
\end{figure*}

We show in this section the whole set of images for the epochs with a detection of the source.  In Fig.~\ref{f.vlbi_in} we show eleven 160 mas $\times$ 160 mas panels illustrating the evolution of the inner ejecta.  The epoch, array, and frequency of each observation is indicated in the panel. In Fig.~\ref{f.vlbi_out} we show 420 mas $\times$ 420 mas panels illustrating the three epochs in which there is emission detected also in the distant component ``K'', in addition to the inner ejecta; all these observations were obtained with the EVN at 1.6 GHz.

\section{An alternative scenario for the motion of the ejecta}\label{s.appendixMotion}

\begin{figure}
\center
\includegraphics[width=\columnwidth]{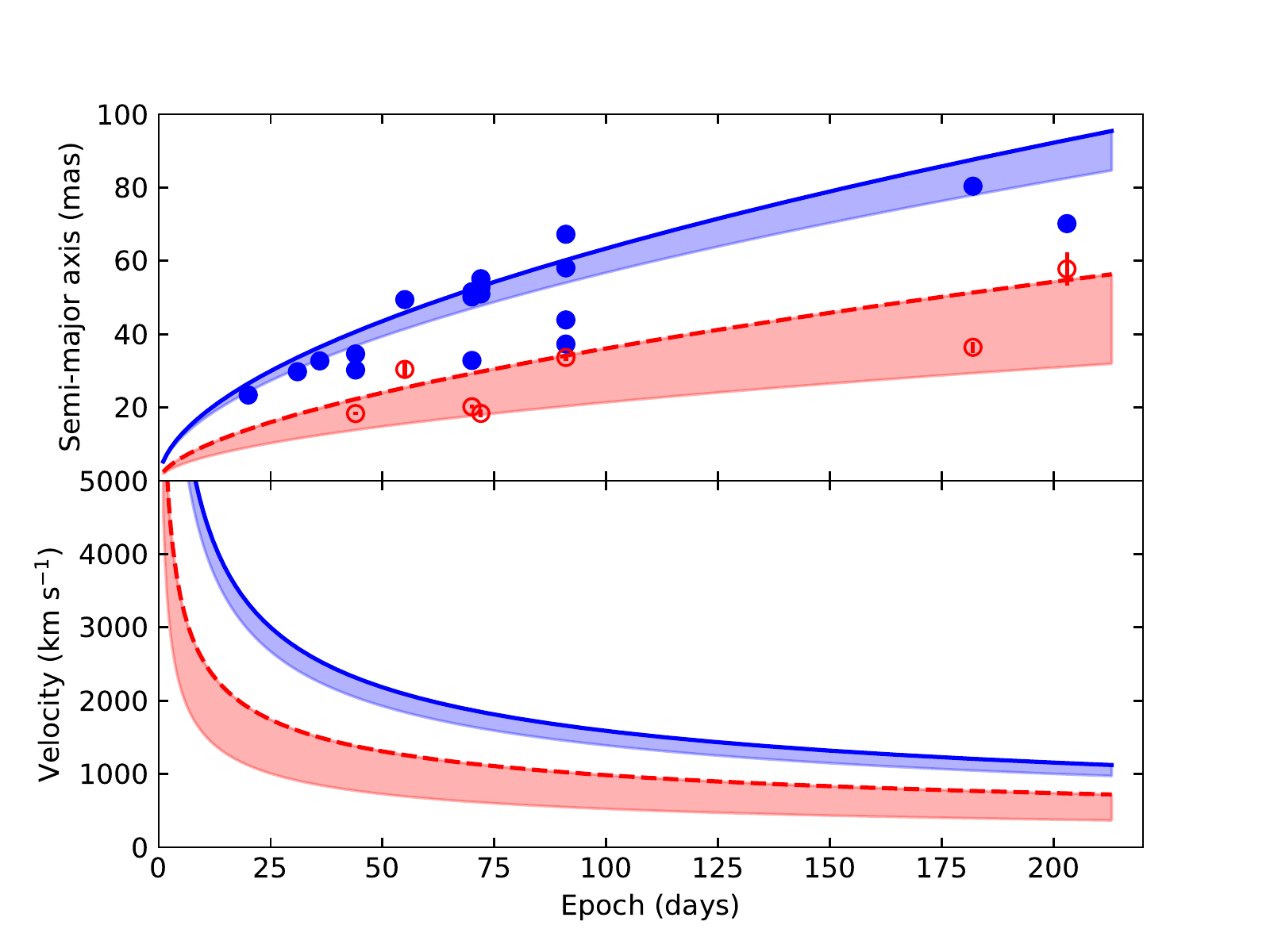}
\caption{Top panel: radial distance of VLBI components from the Gaia position referenced to the nova epoch; filled blue and empty red dots indicate components in the SE and NW regions, respectively; the shaded regions show the best-fit and uncertainty of a $r\propto t^p$ fit to the trend. Bottom panel: linear velocity obtained by deriving the power-law motion, in blue and red for the two sides, respectively.  \label{f.motion_refgaia}
}
\end{figure}

In the main text, we describe the motion of the bipolar ejecta of \vv\ as regions of a growing eccentric ellipse of given centre and axial ratio.  The centre of that ellipse was determined somewhat arbitrarily and it lies within 16.7 mas ($=45$ AU), in p.a.\ 118$^\circ$ degrees, from the Gaia position.  This distance is comparable with the orbital separation of the system.  However, the availability of the Gaia position referenced to the epoch of the nova offers a privileged reference that is worth exploring.  Moreover, the spectral asymmetries between the two sides of the inner structure (see Table~\ref{t.spectrum}) suggest that it might also be worth exploring a scenario in which the two sides do not necessarily share the same physical conditions (both intrinsic or in terms of the surrounding medium).  We have therefore re-referenced the coordinates of every component to the Gaia position, and fit separately the increase of their radii on the NW and SE sides.   For most components, the actual radius in the sky plane could represent an underestimate of the real separation from the origin (if there is a significant component along the line of sight).  Therefore, we highlight the upper values of our uncertainty range as a preferred estimate for the motion.  

The results are shown in Fig.~\ref{f.motion_refgaia}.  The best-fit power-law representing the separation $r\propto t^p$ have indexes $p=0.54$ and $p=0.59$ on the SE and NW sides, respectively, i.e.\ slightly shallower than in the ellipse scenario.  The resulting velocity is always higher on the SW-side, initially by a factor $2\times$ and eventually reaching a factor $\sim1.5$.

Also in this case, as in the symmetric scenario, the velocity can not be well constrained at initial epochs.  We provide reference values of $v_{10}=4585$ \kms, $v_{25}=3010$ \kms, $v_{100}=1590$ \kms, $v_{200}=1160$ \kms, for the SE side, and of $v_{10}=2550$ \kms, $v_{25}=1750$ \kms, $v_{100}=980$ \kms, $v_{200}=740$ \kms, for the NW side (the subscript indicate the epoch, in days).

\end{appendix}

\end{document}